\documentclass[preprint,12pt]{elsarticle}

\usepackage{graphicx}
\usepackage{natbib}
\usepackage{colortbl}
\definecolor{myblue}{rgb}{0.8,0.85,1}

\usepackage{amssymb}
\usepackage{multirow}
\usepackage{float}
\usepackage{subfig}

\usepackage{lineno}

\journal{Nuclear Instruments and Methods A}

\begin{document}

\begin{frontmatter}



\title{A Segmented, Enriched N-type Germanium Detector for Neutrinoless
Double Beta-Decay Experiments
}


\author[add1,add2]{L. E. Leviner}
\ead{lelevine@ncsu.edu}
\author[add5]{C. E. Aalseth}
\author[add3,add2]{M. W. Ahmed}
\author[add5,add9]{F. T. Avignone III}
\author[add1,add2,add17]{H. O. Back}
\ead{hoback@ncsu.edu}
\author[add13]{A. S. Barabash}
\author[add6]{M. Boswell}
\author[add2,add3]{L. De Braeckeleer}
\author[add14]{V. B. Brudanin}
\author[add7]{Y-D Chan}
\author[add14]{V. G. Egorov}
\author[add6]{S. R. Elliott}
\author[add6,add7]{V. M. Gehman}
\author[add5]{T. W. Hossbach}
\author[add1,add2,add5]{J. D. Kephart}
\author[add3,add2,add15]{M. F. Kidd}
\author[add13]{S. I. Konovalov}
\author[add7]{K. T. Lesko}
\author[add12]{Jingyi Li}
\author[add6,add16]{D-M Mei}
\author[add12]{S. Mikhailov}
\author[add5]{H. Miley}
\author[add8]{D. C. Radford}
\author[add11]{J. Reeves}
\author[add14]{V. G. Sandukovsky}
\author[add13]{V. I. Umatov}
\author[add10]{T. A. Underwood}
\author[add3,add2]{W. Tornow}
\author[add12]{Y. K. Wu}
\author[add1,add2]{A. R. Young}
\ead{aryoung@ncsu.edu}

\address[add1]{Department of Physics, North Carolina State University, Raleigh, NC,
USA}
\address[add2]{Triangle Universities Nuclear Laboratory, Durham, NC, USA}
\address[add3]{Department of Physics, Duke University, Durham, NC, USA}
\address[add4]{Department of Physics, University of North Carolina, Chapel Hill, NC,
USA}
\address[add5]{Pacific Northwest National Laboratory, Richland, WA, USA}
\address[add6]{Los Alamos National Laboratory, Los Alamos, NM, USA}
\address[add7]{Lawrence Berkeley National Laboratory, Berkeley, CA, USA}
\address[add8]{Oak Ridge National Laboratory, Oak Ridge, TN, USA}
\address[add9]{Department of Physics, University of South Carolina, Columbia, SC, USA}
\address[add10]{AMETEK/ORTEC, Oak Ridge, TN, USA}
\address[add11]{Reeves and Sons, LLC, Richland, WA, USA}
\address[add12]{Duke University Free-Electron Laser Laboratory, Durham, NC, USA}
\address[add13]{Institute of Theoretical and Experimental Physics, Bol'shaya Cheremushkinskaya ul. 25, Moscow, 117259 Russia}
\address[add14]{Joint Institute for Nuclear Research, Dubna, Moscow oblast, 141980 Russia}
\address[add15]{Tennessee Tech University, Cookeville, TN, USA}
\address[add16]{University of South Dakota, Vermillion, SD, USA}
\address[add17]{Princeton University, Princeton, NJ, USA}

\cortext[cor1]{L. E. Leviner}

\begin{abstract}
We present data characterizing the performance of the first
segmented, N-type Ge detector, isotopically enriched
to 85\% $^{76}$Ge.  This detector, based on the Ortec PT6x2 design and
referred to as SEGA (Segmented, Enriched Germanium Assembly), was
developed as a possible prototype for neutrinoless double beta-decay
measurements by the {\sc Majorana} collaboration. We present some of the
general characteristics  (including bias potential, efficiency,
leakage current, and integral cross-talk) for this detector in its
temporary cryostat.  We also present an analysis of the resolution of
the detector, and demonstrate that for all but two segments there is at least one channel that reaches
the {\sc Majorana} resolution goal below 4 keV FWHM at 2039 keV, and all channels are below 4.5 keV FWHM.

\end{abstract}


\end{frontmatter}


\section{Introduction}
\label{}
The Segmented Enriched Germanium Assembly (SEGA) is a
prototype detector for the {\sc Majorana} experiment that is predominately focused on the
search for neutrinoless double beta-decay of  $^{76}$Ge \cite{Aalseth:2004p148,Schubert:2012p480,Wilkerson:2012}.  The
observation of neutrinoless  double beta-decay would establish that
neutrinos are Majorana particles, show that lepton number is not conserved, and
provide information concerning the absolute neutrino mass scale \cite{Camilleri:2008p343,III:2008p481,Barabash:2010p162}.  The ultimate goal of the 
{\sc Majorana} experiment is to probe neutrinoless double beta-decay
half-lives greater than 10$^{27}$ years and the
absolute neutrino mass scale at 20-40 meV.

The search for such rare events requires significant effort to reduce the effects of
background events, which include cosmic rays, decay events from
natural and cosmogenic radioisotopes, and two-neutrino double beta-decay
events.  Methods to eliminate backgrounds include the use of ultra-pure
materials, shielding from external sources, optimization of energy
resolution, segmentation, and pulse-shape analysis (PSA).  The goal of
the tonne-scale {\sc Majorana} experiment is to reduce background events to 1
count/tonne/year in an approximately 4 keV region-of-interest (ROI) centered around
the $^{76}$Ge Q-value of 2039 keV.  The {\sc Majorana} collaboration is
currently focusing development efforts towards P-type, point contact
detectors \cite{Barbeau:2007p53} for the {\sc Majorana Demonstrator} project \cite{Schubert:2012p480,Wilkerson:2012,Phillips:2012};
however the remarkable event reconstruction and
background rejection capabilities of segmented N-type detectors make
them an interesting alternative \cite{Elliott:2006,Deleplanque:1999p292,Abt:2007p479}.
For both of these geometries, optimum background rejection is achieved by
analysis of pulse shapes to differentiate backgrounds from a potential neutrinoless double beta-decay signal.  
This is possible primarily because double beta-decay gives rise to a ``single-site'' event in 
which ionization is deposited in a very small volume, with a maximum linear dimension of about a millimeter,
whereas events originating from background gamma-rays typically produce multiple Compton scattering sites typically separated
by a few centimeters.  By analyzing the pulse-shape in each physical segment of our N-type detector, one can 
reconstruct and reject events which produce ionization in more than one segment or result in multiple ionization events 
in a single segment.  A detailed analysis of the pulse-shapes and background rejection capabilities is 
the subject of an article in preparation \cite{SEGAPSA}.  For more details on background rejection strategies, refer to references \cite{Aalseth:2004p148,Deleplanque:1999p292,Abt:2007p479}.

The SEGA detector is the first $^{76}$Ge enriched, segmented, N-type High Purity Germanium (HPGe)
detector.  The GRETINA \cite{Descovich:2005p109}, AGATA \cite{Recchia:2009p111}, and GERDA \cite{Abt:2007p474,Abt:2009} collaborations have already
demonstrated the capability of reconstructing the event topology for
high-energy (near 2 MeV) events associated with double beta-decay
with position resolution of a few millimeters through analysis of pulse shape and the
image-charge distribution in similar, highly segmented N-type
detectors.  SEGA has also served as a test bench for
engineering low-background crystal mounts and cryostats components for
the N-type segmented geometry. Finally, after installation in a low
background cryostat, SEGA should also be a valuable tool for a variety of background studies.

In this article, we report on the general characteristics and performance of
the SEGA detector in a temporary cryostat.  The fragility of the
outer, P-type lithographic contacts presents some risk in the transfer
to a permanent, low background set of contacts, so we present the
preliminary performance of this system as a benchmark in our progress
towards implementing a low background configuration underground. 
The electronic noise characteristics were not optimal, which was primarily due to 
``warm" preamplifiers, with long contact-to-preamplifier wires.
However, our analysis of the system indicates that SEGA can meet the
performance specifications for a large scale double beta-decay
experiment when equipped with cold FETs, mounted as close as possible
to the crystal.

\section{The Detector}
\label{}

\begin{figure}[h!]
 \centering \includegraphics[width=1.0\textwidth]{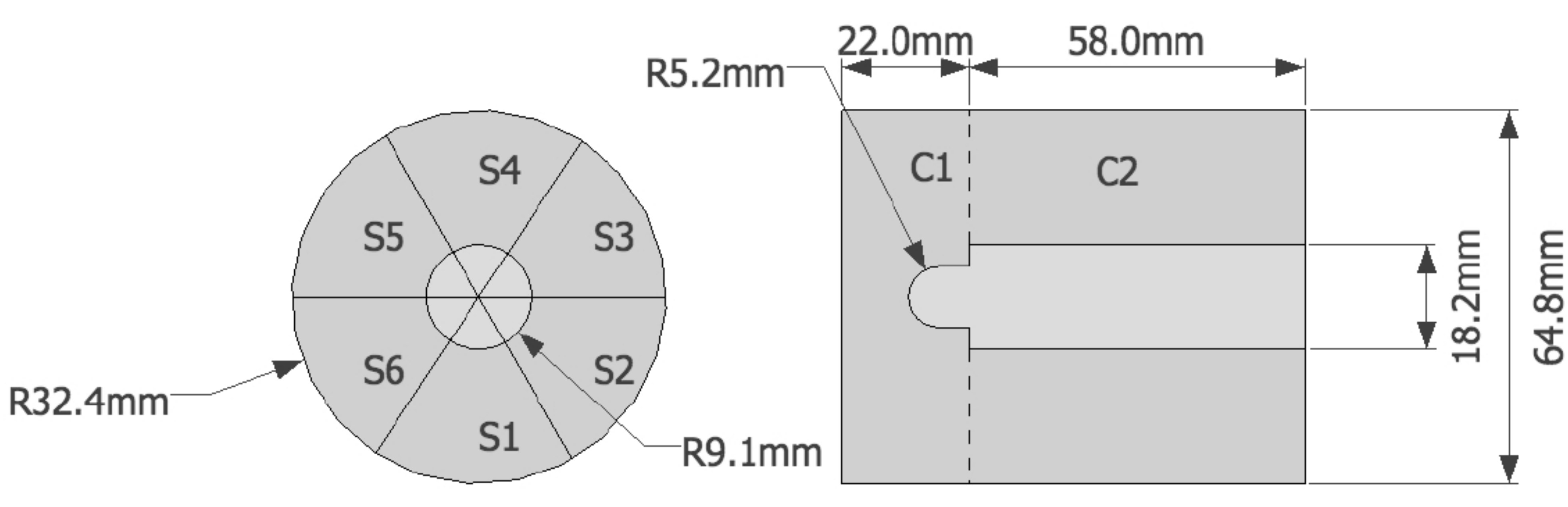}
 \caption{The ORTEC PT6x2 detector geometry.  The SEGA detector is not
``bulletized", which means there is no radius for the closed-end,
 outer edge.}
  \label{fig:pt62}
\end{figure}

SEGA is an N-type HPGe detector enriched to 85\% $^{76}$Ge with 6 outer,
azimuthal and 2 central, axial segmentations.  The Ge material used to produce SEGA was composed 
of three separate batches from ITEP \cite{itep:comp}, IGEX \cite{igex:comp}, and Dubna \cite{dubna:comp}, with the respective $^{76}$Ge enrichment abundances 
and masses of 87.1\% [4 kg], 80.8\% [3.7 kg], and 86.8\% [3.3 kg].  
The 1.374 kg detector has a diameter of 64.8 mm and a total length 
of 80.0 mm.  The six outer contacts
were produced by implanting the masked Ge with 100 keV $^{11}$B ions.  The resulting 
inter-segment gap is 170 microns.  The two internal lithium contacts are
produced using novel, monolithic segmentation technology developed by ORTEC \cite{ortec:comp}.  
The internal contact segmentation results in a (C2) contact which is situated on the surface of a
18.2 mm diameter hole bored 58.0 mm into the crystal and a (C1) contact positioned on a 10.5 mm dimple at the bottom of the bored hole.  This segmentation
geometry results in 12 physical segments (six in a cylindrical
geometry and six in a ``hockey puck" geometry) and is referred to as
the ORTEC PT6x2 geometry  \cite{Sangsingkeow:2003p34}, with an idealized  example depicted in
Fig. \ref{fig:pt62}.  

A possible advantage of this scheme for double beta-decay measurements 
is that the resultant 12 physical segments are produced with only eight instrumented 
contacts, minimizing the required front-end electronics per physical segment\footnote{The advantage
of fewer contacts per physical segment will be partially offset by the activity of the components required for
the central contacts inserted into the inner bore of the crystal.}.   
We note that the resultant segment volumes were near those selected for 
prototype designs of the {\sc Majorana} experiment using N-type detector arrays \cite{Aalseth:2004p148}.

The labeling of the twelve physical segments of the SEGA detector
follows a CXSY format, where X and Y represent one of the two central
and six outer contacts, which may also be referred to as channels,
respectively.  In order to avoid loss of enriched Ge in this prototype detector,
the closed end, outer surface of the SEGA detector was not ``bulletized", as indicated
in Fig. \ref{fig:pt62}, resulting in low electric fields and poorer charge collection
in the corners of the closed end.  Detectors used in a germanium double beta-decay experiment would be bulletized.

\section{Test Environment}
\label{}

\subsection{Temporary Cryostat}

All of the tests and performance data presented in this document were
taken with the SEGA detector mounted in a temporary cryostat. The
detector was mounted vertically with the central bore oriented with
the hole upward (C2 above C1). The outer contacts are copper
leaf-springs with a 25 mm$^{2}$, 0.5 mm thick indium foil at the point
of contact between the spring and the detector face. The central
contacts were mounted to a brass rod which was inserted into the central
bore of the detector: the C2 contacts were leaf-springs attached to
the rod and the C1 contact was a gold-coated brass plug mounted on the
end of the rod (with Indium foil at the point of contact with the
crystal) and kept in contact with the crystal. The detector
was mounted in a central shield to provide thermal isolation and held
in place with PTFE pads (with Cu coated
PTFE straps to provide good thermal contact with
the base plate).  The cryostat was configured as a dip-stick and inserted
in a standard, ORTEC LN$_{2}$ dewar. The temperature of the crystal
was maintained  at 93 to 97 K, as monitored by a platinum resistor
mounted to the base plate.

The detector signals were routed out of the cryostat to HeKo preamplifiers \cite{Eberth:2001p228,Gamir:1997p113,canberra:comp}
with warm FETs, which have an input capacitance of 15 pF 
and a maximum energy throughput of about 10 GeV/s.
When the detector is biased, the inner segments have a calculated
capacitance of 14-20 pF for C1 and 40 pF for C2, and the outer segments
have calculated capacitances of 9-10 pF.  The high voltage is applied to the 
central contacts via separate 2 G$\Omega$ load resistors with the two central contacts
being AC coupled to the preamps via 2.7 nF, 6 kV ceramic capacitors.  The
signals were processed using two XIA DGF-4C \cite{xia:comp} (Digital Gamma Finder with
4 Channels) modules, which digitized the signals using 14-bit, 40 MHz
ADCs.  These modules can store up to 100 $\mu$s waveforms per event,
perform pulse-shape analysis, and extract a real time energy using an
on-board trapezoidal filter into a spectrum with up to 32K channels.
Each channel has a real-time processing unit (RTPU), which is equipped with 
a field programmable gate array (FPGA) and a FIFO memory.  When an event is 
detected in the FPGA, a trigger is issued to the digital signal processor (DSP), which
then processes the data and prepares it to be read out by the CAMAC interface to 
the computer.  The system is configured to read out all eight channels
when a trigger is issued in either of the 2 central-contact channels.
The configuration of the system in the temporary cryostat did not accommodate instrumenting
a pulser for all contacts, so the response of the detector system to
external radioactive sources was used to characterize the detector
performance.

\section{Detector Properties}
\label{}

\subsection{Electric-Field Profile}

\begin{figure}[H]
 \centering \includegraphics[width=0.750\textwidth]{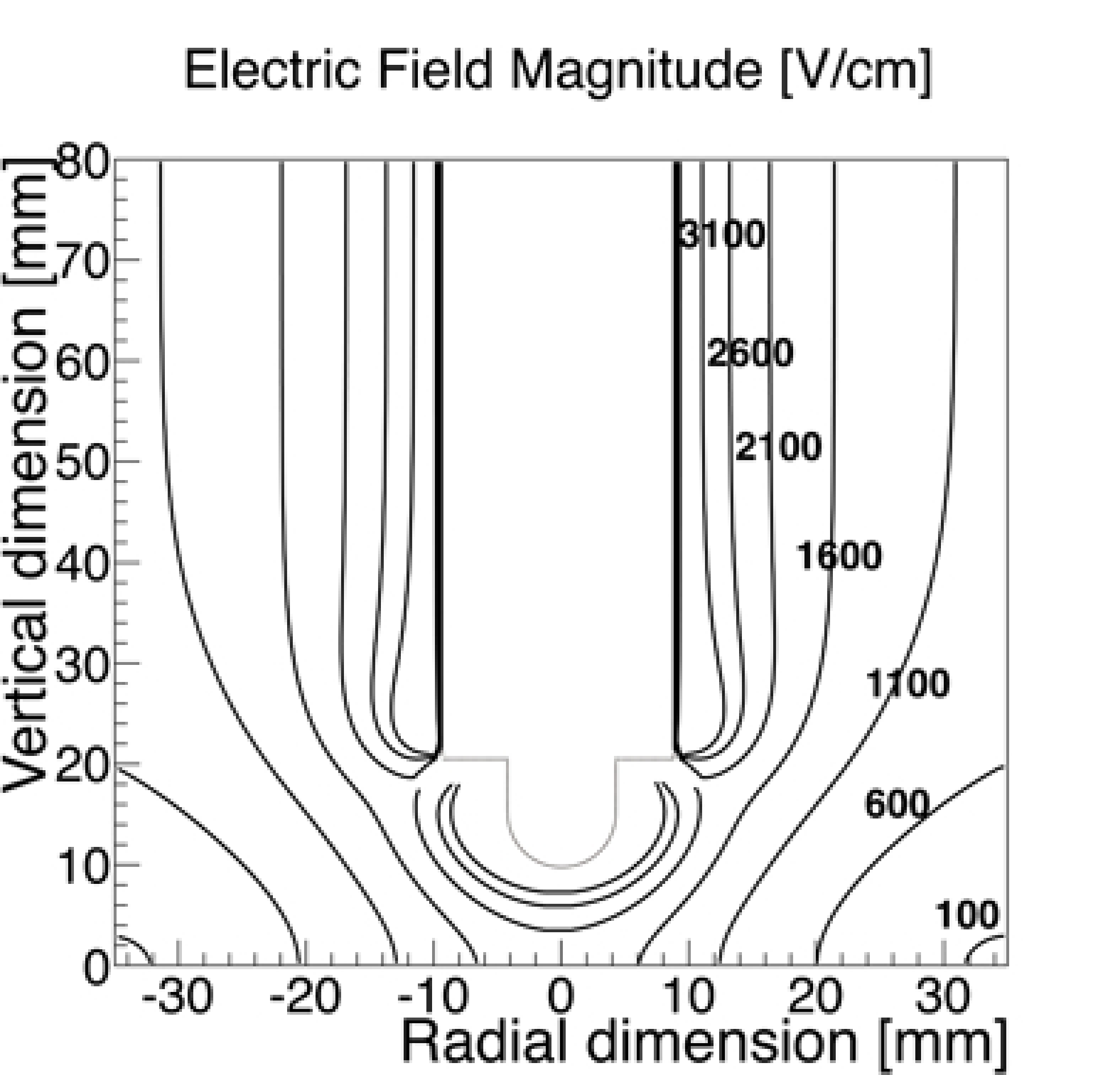}
 \caption{Calculated electric-field profile in the SEGA crystal.  The inner electrode details are not shown per vendor request, hence discontinuities are present
 for some of the field lines.}
 \label{fig:fieldprofile}
\end{figure}

For Pulse-shape Analysis (PSA) studies, the electric-field profile for SEGA was calculated (see Fig.
\ref{fig:fieldprofile}). An impurity concentration, based on measurements at ORTEC,  which varied from
3.0$\times 10^{9}$/cm$^{3}$ at the open end of the crystal to
2.7$\times 10^{10}$/cm$^{3}$ at the closed end was used, along with
no radial gradient.  From this electric-field profile it is evident
that the closed end has low-field regions near the edge of the
crystal where charge collection is expected to be less efficient.

In efforts to quantitatively assess the effects of the low-field regions, rise-time
and low-energy tail analyses were done with a $^{60}$Co source.  It was 
found that C1 segment's rise-time means were greater than C2 segment's rise-time means, with all 
segments' means ranging from 170 ns to 250 ns and less than 0.2\% events having rise-times greater than 400 ns.  
Low-energy tails on the photopeaks were also found to be present and the degree of low-energy 
tailing is shown in Figs. \ref{fig:lowetail} and \ref{tab:fw2} (See Appendix C).  The 1332 keV photopeaks
were fit with both a Gaussian and a modified Gaussian function \cite{Debertin:1988p164} shown below, which took into account the low-energy tail.  
\begin{equation}
a_0\mathrm{e}^{\frac{-(x-a_1)^2}{2{a_2}^2}} + a_3\mathrm{e}^{\frac{-(x - a_1 + a_4)^2}{(2{a_2}^2a_5)}} + a_6(\mathrm{Erf}(a_7(x-a_1)) + 1)
\end{equation}
As expected, due to the low-field regions, the FWFM/FWHM ratios in
Table \ref{tab:fw2} show significant deviations of the modified Gaussian
functional fit from the Gaussian functional fit for the SXC1 channels.  To further support the deviations
from a Gaussian functional fit, the integrated number of counts in the low-energy tail as a ratio to the number of counts in the photopeak is presented in Table \ref{tab:fw2}.
The increased size of the low-energy tails for the S channels coupled to C1 as opposed to C2 is evident.

\begin{figure}[H]
 \centering \includegraphics[width=0.750\textwidth]{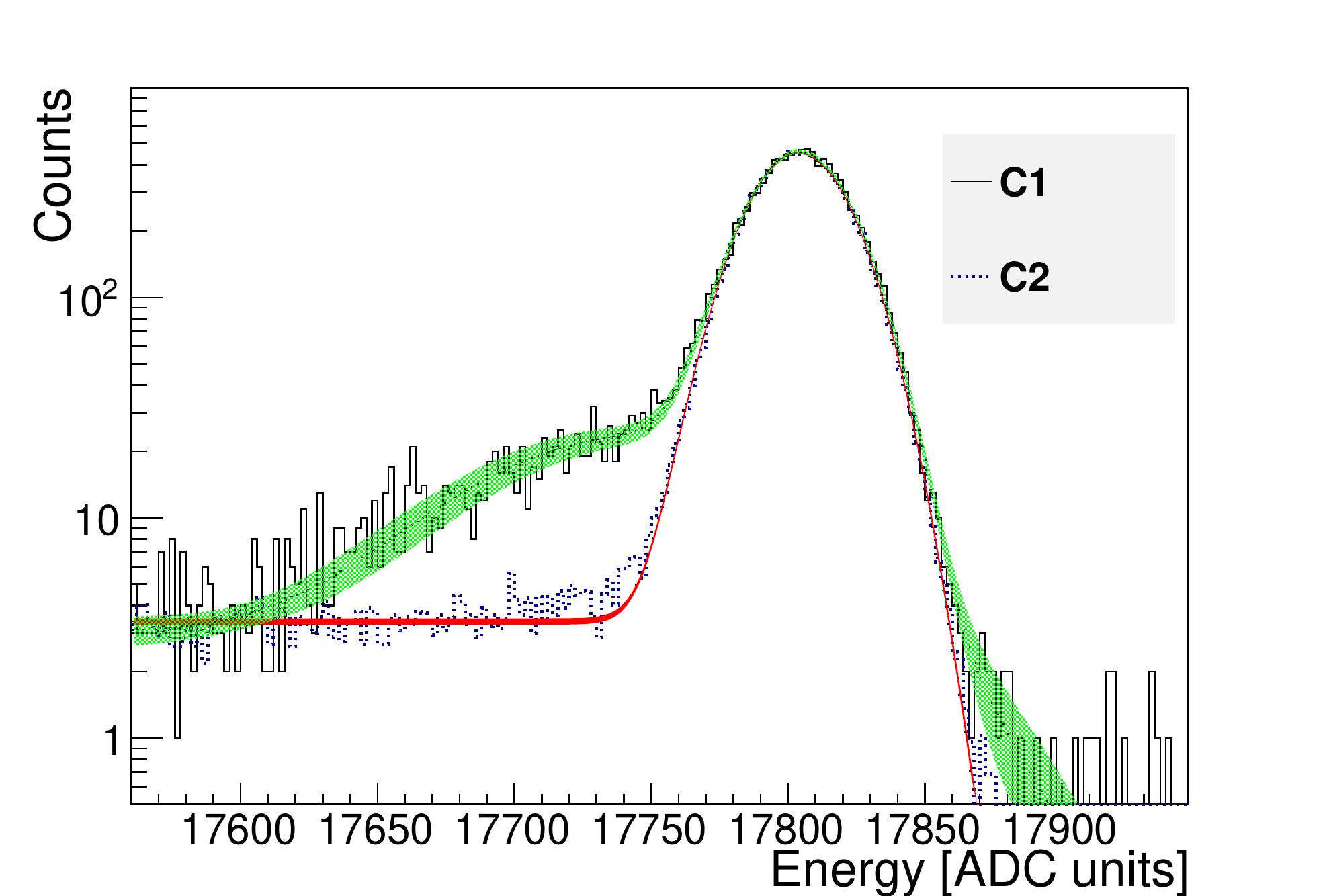}
 \caption{1332 keV photopeak for channel S1 with S1C1 and S1C2 single segment cuts.  The degree of low-energy tailing is clearly greater in the C1 segment
 where the low-field regions are present.  The width of the fit band reflects uncertainties of the fits to the low-energy tail.}
 \label{fig:lowetail}
\end{figure}

\subsection{Depletion Voltage}

\begin{figure}[H]
 \centering \includegraphics[width=0.75\textwidth, angle=0]{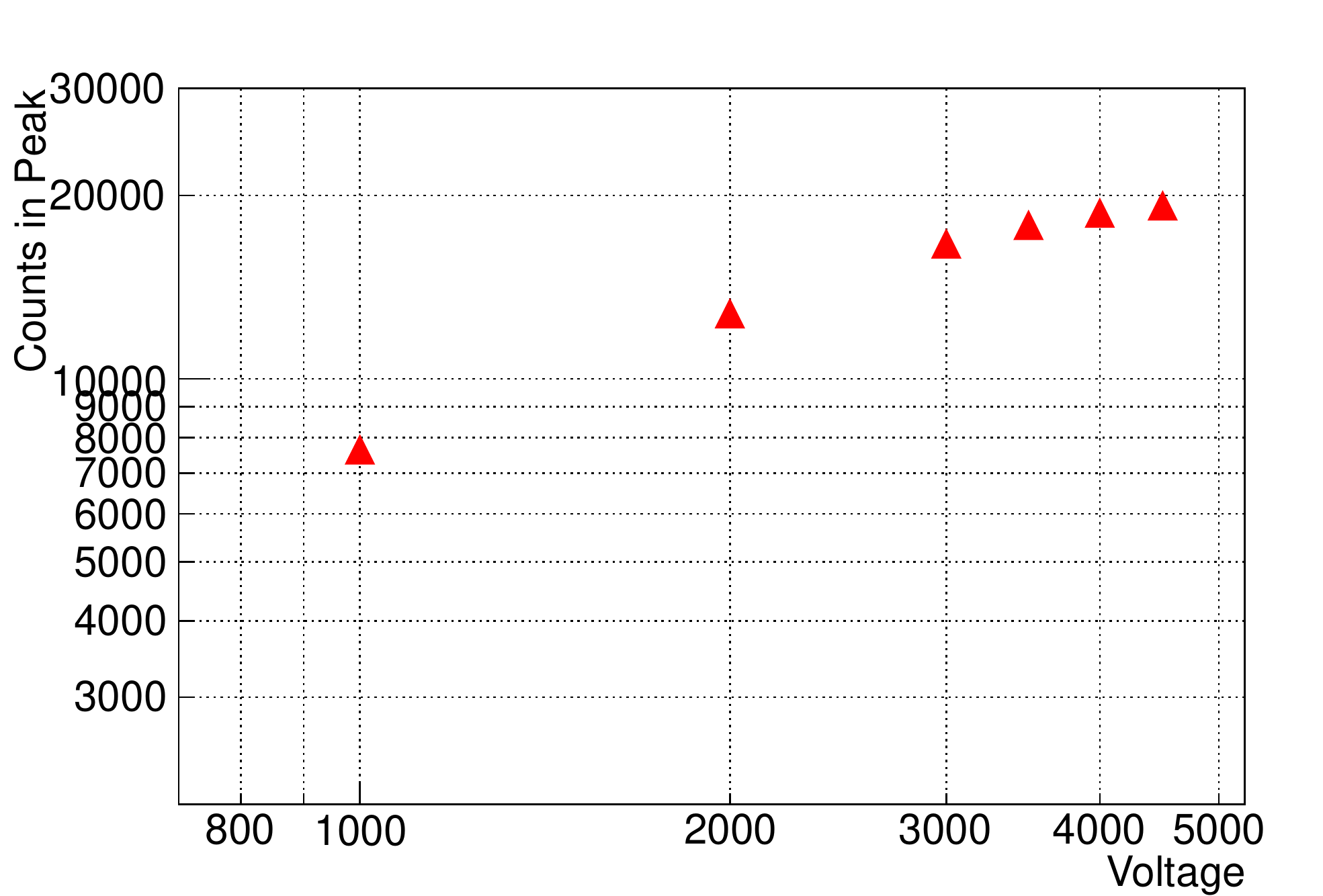}
 \caption{Counts in the $^{60}$Co 1332 keV $\gamma$-ray peak (for the C1
 and C2 channels connected together) as a function of bias voltage.  The slope of the efficiency curve approaches zero as the operating voltage is reached, confirming that effectively all charge deposited in the detector
 is collected.}
 \label{fig:fullpeakenergy}
\end{figure}

The depletion voltage was determined at ORTEC by peak efficiency
measurements of the 1332 keV $^{60}$Co $\gamma$ rays as a function of bias
voltage while connecting the two central contacts (C1 and
C2). From these measurements and observations of the behavior of the
central contacts which become independent only when the
detector is almost fully depleted, it was determined that the depletion
voltage for the crystal is 4750 V and an operating voltage of 4800 V
was deemed appropriate (see Fig. \ref{fig:fullpeakenergy}).

\subsection{Leakage Current}

One challenge associated with the fabrication of SEGA was the
reduction of the leakage current to an acceptable level.  In order to
accomplish this task without reprocessing all of the enriched germanium
multiple times, ORTEC re-etched the surface of the crystal
and produced new contacts until an  ``acceptably" low leakage current was
obtained.  Ultimately the leakage  current was determined by ORTEC to be
roughly 350 pA by connecting together all of the outer contacts (S1-S6)
and measuring the current to ground with a bias voltage of 4800 V
applied to the central contacts C1 and C2.  Even so, this leakage
current was about factor of two above what ORTEC considered  ``optimal".  It should be noted that, ORTEC also performed similar leakage 
current measurements of a non-enriched prototype detector with the PT6x2 geometry and found
the leakage current to be 72 pA and 80 pA at 3200 V and 3300 V respectively \cite{Sangsingkeow:2003p34,ORTEC:private}\footnote{3200 V was the operating voltage of the non-enriched prototype PT6x2 detector.}.
For a quantitative treatment of the impact of leakage current on the resolution
see Sec. 4.7 and Appendix B.

\begin{figure}[H]
 \centering \includegraphics[width=0.75\textwidth, angle=0]{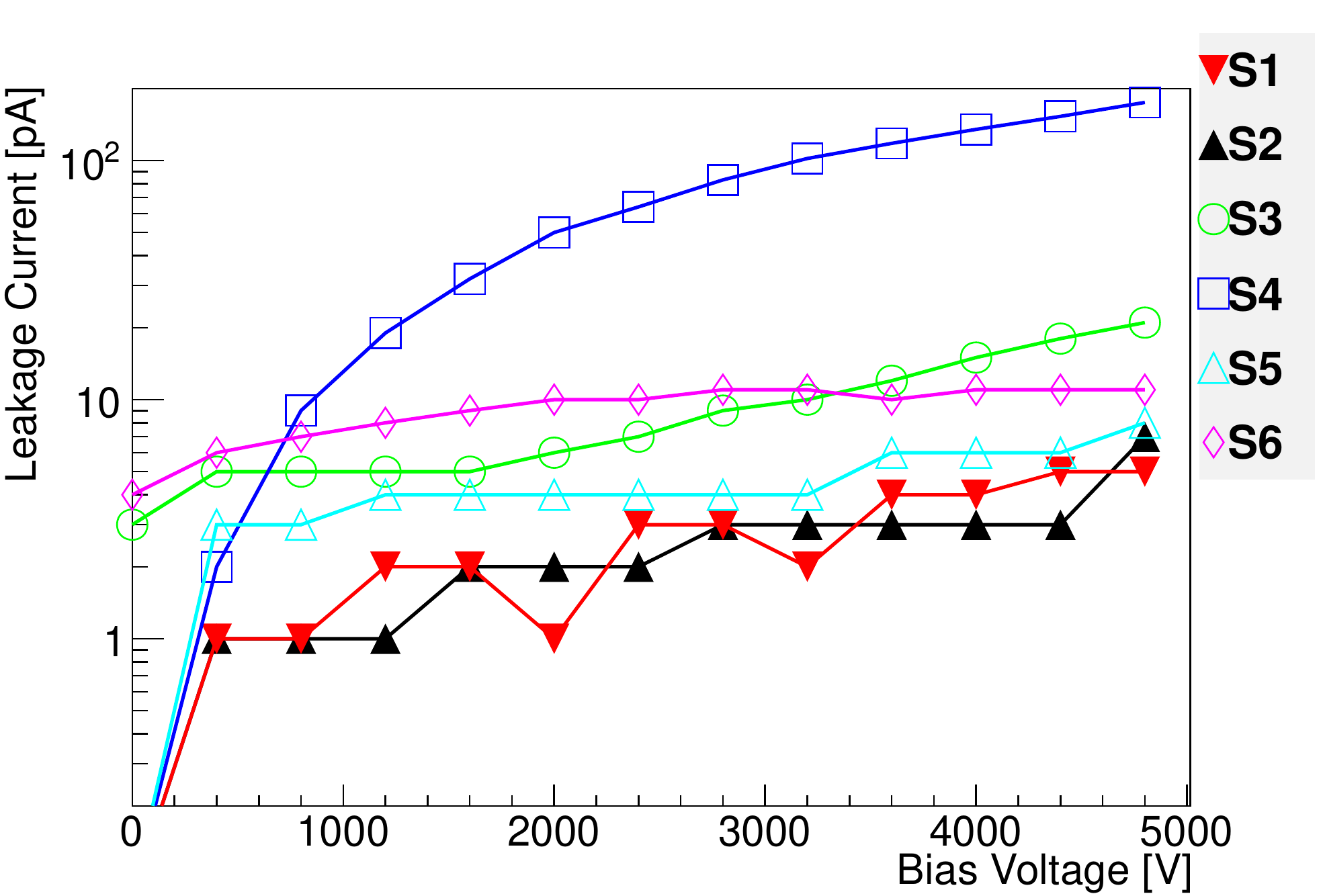}
 \caption{Leakage current from segments S1 to S6 as a function of bias applied to
 C1 and C2.  These measurements were done in February, 2003.}
 \label{fig:leakage}
\end{figure}

Additional measurements of the leakage current for each segment were
performed at Triangle Universities Nuclear Laboratory (TUNL) soon after receiving the 
SEGA detector from ORTEC.  For
these measurements the detector was fully biased and left overnight,
since it was found the leakage current would decrease significantly
after the first few hours after applying bias to the detector.  The
following morning, the voltage was slowly  stepped down and the
leakage current measured through each outer segment 
after the current was allowed to settle for
approximately 15 minutes.  From this study it was found that the dominant 
source of leakage current was segment S4 (see Fig. \ref{fig:leakage}).  This overall pattern (segment S4 being the dominant source of leakage current) was
maintained over the 10 year period for which the measurements documented in this work were performed.  However, there was a significant
increase in the magnitude of the leakage current over this period (see Sec. 4.7 and Appendix B for details).

\subsection{Efficiency}

In order to assess the active volume within the detector, the intrinsic peak
efficiency was measured using a 3.15 $\mu$Ci $^{60}$Co source.  First, by scanning with a 
$^{57}$Co source it was determined that the crystal, within the cryostat, was positioned 4.1 cm from the endcap.  Then, the $^{60}$Co source was placed
approximately 25 cm above the top of the crystal.  It should be noted that this method is slightly different from the IEEE standard efficiency measurement \cite{Keyser:eff} where the 
source is placed 25 cm from the endcap.  Data were recorded for approximately 43 hrs to
reduce the statistical uncertainty in the sum-peak
to less than 5\%.  The rate was approximately 700 Hz
with our system livetime near 20\%, which was limited by the throughput of the CAMAC controller.  
For these data the trapezoidal filter rise and gap times 
were set to 10.8 $\mu$s and 1.6 $\mu$s respectively (corresponding to an effective shaping time of approximately
6 $\mu$s  for semi-Gaussian filters) \cite{xia:comp}.

\begin{figure}[H]
\begin{center}
\includegraphics[width=0.7\textwidth]{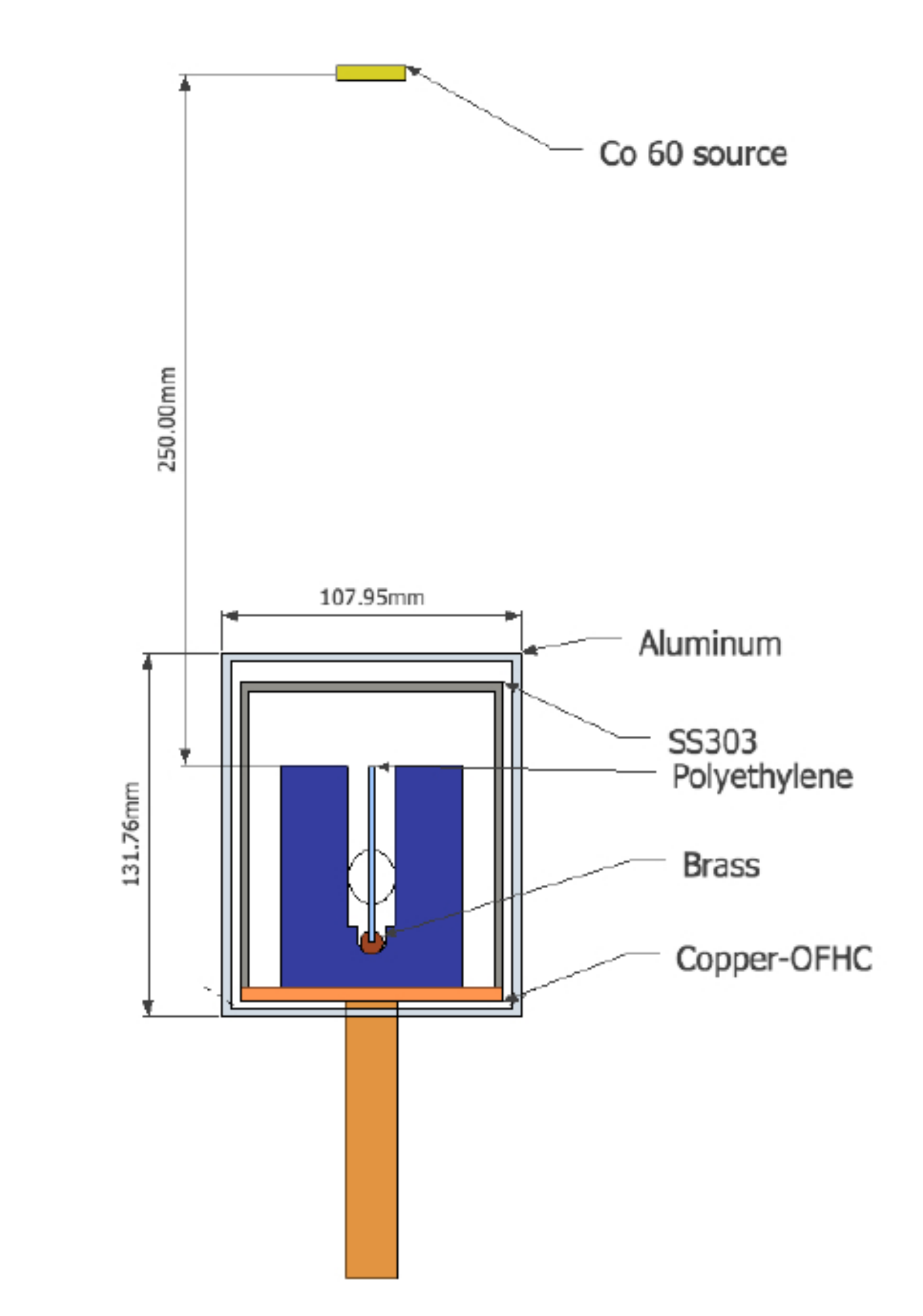}
\caption{
SEGA detector geometry used in MaGe simulation.  The cold finger and some small parts were not simulated.}
\label{fig:simgeo}
\end{center}
\end{figure}

To determine the full energy
deposited within the detector for a given event, the energy for each of the outer S channels (S1-S6)
were summed, and because of the coupling between segments, a cross-talk correction \cite{Bruyneel:2009p93} was applied (see Fig. \ref{fig:xtalkcomp} in Sec. 4.6).
By analyzing the number of counts in the two $^{60}$Co $\gamma$
peaks and in their sum-energy peak (2505 keV), it was possible to determine the intrinsic peak efficiency at 1173 keV
and  1332 keV, which were 13.9$\pm$1.7\% and 12.8$\pm$1.3\% respectively, where the error value is dominated
by a 5\% uncertainty in the source and detector relative positioning.  Sum-coincidence 
and angular correlation effects were accounted for in this study, while the effect
of system dead time was avoided using the sum-peak method.  These measurements were 
found to be in reasonable agreement with a simulation of the as built SEGA detector geometry 
using the MaGe package \cite{Boswell:2011p1212}, which yielded values of 14.9$\pm$0.3\% and 13.8$\pm$0.2\% respectively.  
The simulated geometry is depicted in Fig. \ref{fig:simgeo}.  For  1332 keV the measured and simulated efficiency values 
relative to a 7.62 cm diameter $\times$ 7.62 cm length NaI(Tl) detector are 108$\pm$11.0\% and 115$\pm$1.7\% respectively.

\subsection{Calibration and detector linearity}

Energy measurements were either made using the trapezoidal filter
in the XIA, DGF-4C module \cite{Tan:2004p22}, or in some special cases, directly from
trace data.  The XIA trapezoidal filter is designed to be relatively
insensitive to the signal rate and uses a large running sample of the
baseline to accurately determine its contributions to the filter
sums. The filter takes as input the preamplifier decay times 
(approximately 48 $\mu$s), which were determined by fitting
large samples of 40 $\mu$s long traces for each segment.

\begin{figure}[H]
\begin{center}
\includegraphics
[width=0.75\textwidth, angle=0]{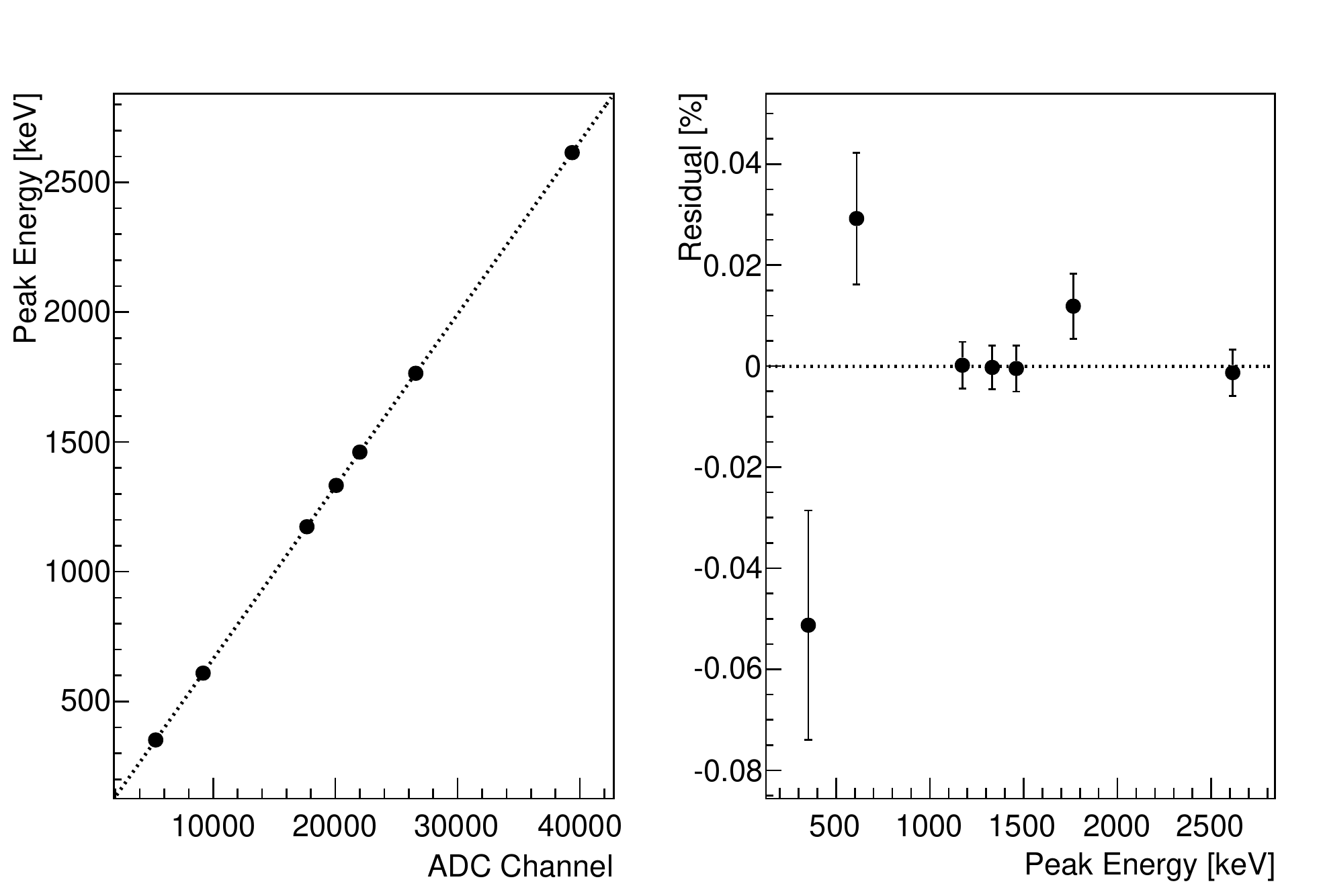}
\caption{
Energy response of C2 channel for C2S2 events from a $^{60}$Co source and various background sources, along with
the residuals of the data-points from the linear fit.}
\label{linearity1}
\end{center}
\end{figure}

The system linearity was
analyzed using the same data set from the efficiency study and fitting to
the various backgound and $\gamma$-ray lines listed in Table \ref{tab:linsource} using a
Gaussian function with a linear background.  The centroids from the Gaussian fits to the peaks
were then fit with a linear function.  The results are depicted
in Fig. \ref{linearity1}, where we note that the y-intercepts were all smaller than 0.56 keV and the integral
linearity was better than 0.067\% at one sigma level for all channels.  

We also note the detector was calibrated
a number of times (over a period of 10 years) using linear and quadratic functions along with a variety of lines.  There was 
almost no variation in the extracted calibration parameters and the quadratic terms were 
less than 4.2$\times 10^{-9}$ keV resulting in corrections of less than 0.2\% to the calibrated energy values out to 2039 keV.

\begin{table}[H] 
\begin{center}
\begin{tabular}{cc}
  \hline
  Source & Energy [keV]  \\ \hline
  $^{214}$Bi & 351.932 \\
  $^{214}$Bi & 609.312 \\ 
  $^{60}$Co & 1173.237 \\ 
  $^{60}$Co & 1332.501 \\ 
  $^{40}$K &  1460.830 \\ 
  $^{214}$Bi & 1764.494 \\ 
  $^{208}$Tl & 2614.533 \\ \hline
\end{tabular}
\end{center}
\caption{$\gamma$-ray sources used for system linearity evaluation. }
\label{tab:linsource}
\end{table}

\subsection{Cross-talk}

Two different methods were used to determine integral cross-talk.  One
required selecting events where all of the $\gamma$-ray energy
deposition is in one physical segment and measuring the shifts in the
measured energy in the non-triggered channels (band method).  Another required the
construction of so-called ``superpulses", which are the average pulses of
many events of the same energy deposition and segment.  For this
analysis both methods were used.

\begin{figure}[H]
\begin{center}
\includegraphics
[width=0.75\textwidth, angle=0]{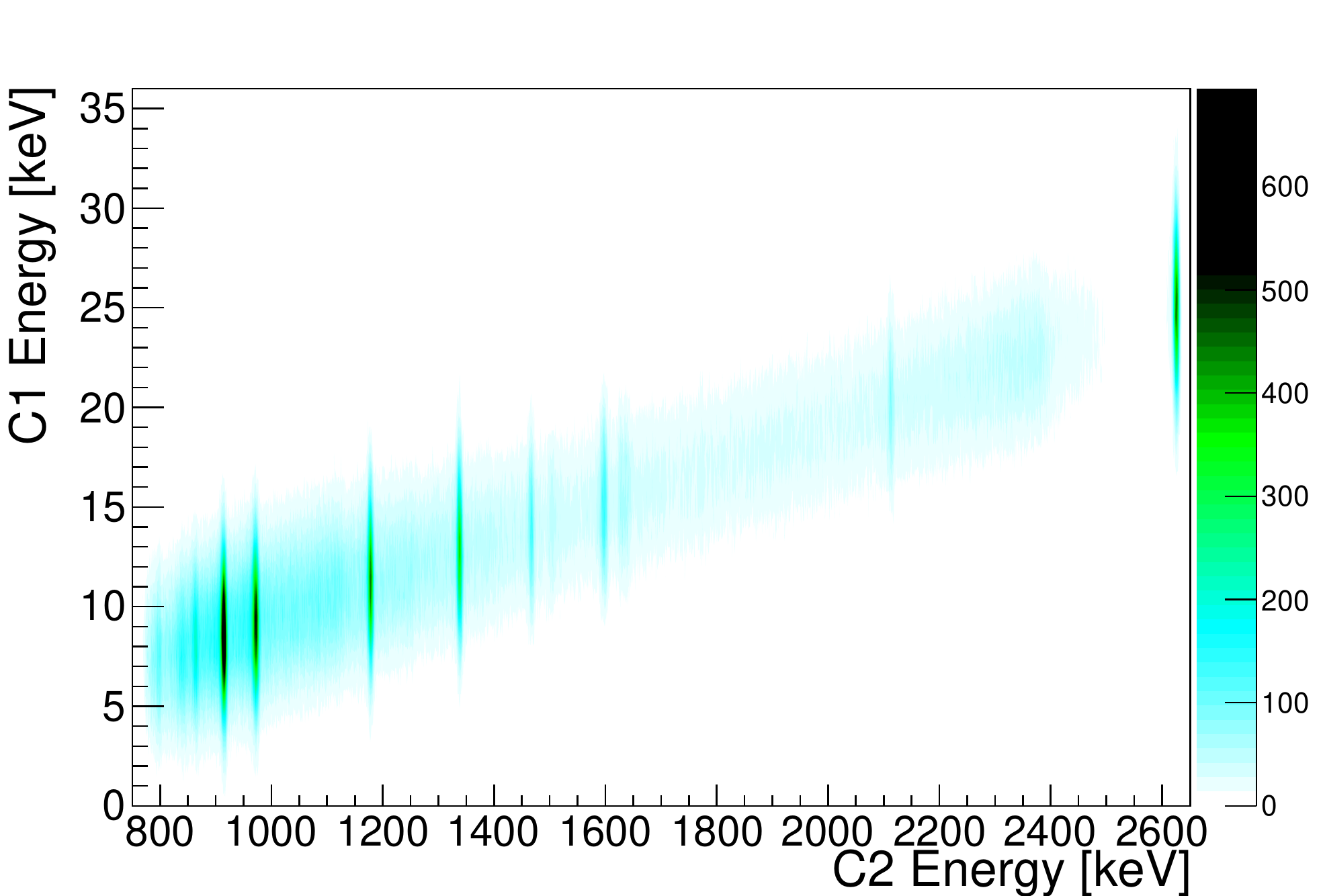}
\caption{Cross-talk band in C1 for C2S2 events.
}
\label{xtalk_cc}
\end{center}
\end{figure}

For the band method it was first necessary to develop single-segment
cuts where only two channels, a central contact and an outer contact,
had an energy value greater than the nominal threshold of about 700 keV.  
The other channels should have a measured energy value of zero, but due to
cross-talk this is not necessarily true.  By performing single-segment
cuts and plotting the triggered channel's energy {\it versus} each
non-triggered channel's energy, cross-talk bands become apparent (see Fig. \ref{xtalk_cc})
and fitting to the slope of the cross-talk bands provides the cross-talk coefficients necessary
to account for this effect \cite{Bruyneel:2009p93}.

A few issues complicated this analysis.  An initial
problem was that the threshold value settings in the XIA DGF-4C module for these data
were around 700 keV and channels which
were not over threshold had an XIA energy filter value of zero.  This
was a problem because energies of just a few keV are characteristic of
cross-talk between contacts. It was also necessary to avoid spurious triggers
and incorrect integrated energy values reported by the XIA DGF-4C module when large image 
charge was present.
To address these issues the trace
from each channel was used to determine the energy using an offline trapezoidal
filter \cite{Tan:2004p22}.  This filter is similar to the trapezoidal filter the XIA DGF-4C
module uses to calculate the energy, but results in lower
resolution because the trace samples used for this particular study were
3 $\mu$s long and included only about 300 ns of baseline data.  The filter resulted in about a factor of two poorer resolution than
the XIA energy filter, but did not impact the cross-talk measurement.

\begin{figure}[H]
\begin{center}
\includegraphics [width=1.0\textwidth, angle=0]{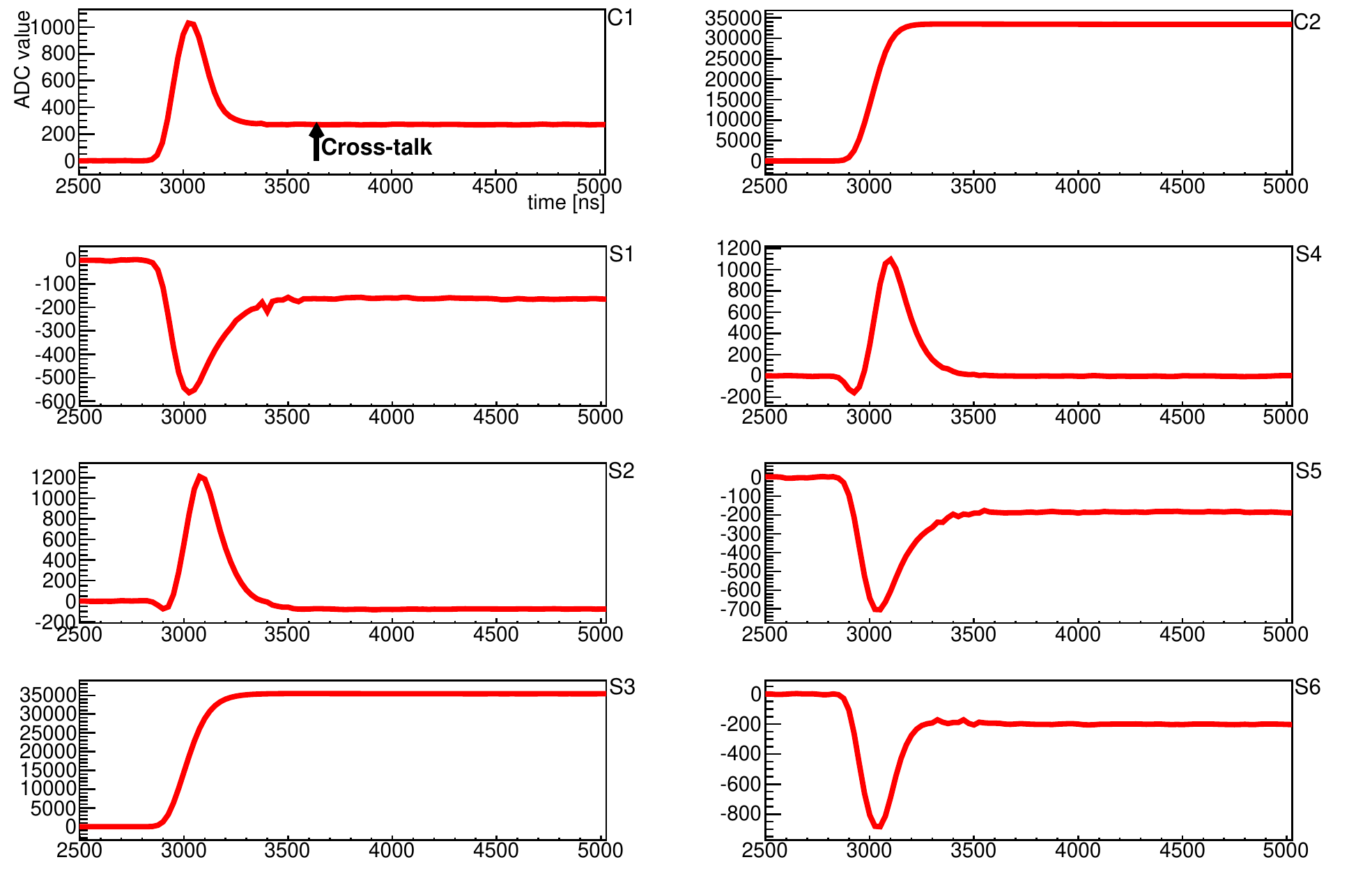}
\caption{Superpulses (ADC values {\it versus} time) from each preamplifier channel for $^{60}$Co
 1332 keV $\gamma$-ray events in segment C2S3.  C2 and S3 show a clear
charge from an event, while the overall non-zero charge in channels
C1, S1, S2, S4, S5, and S6 indicate cross-talk between channels.}
\label{superpfig}
\end{center}
\end{figure}

The second method involved the construction of ``superpulses", which are
constructed for all segments, in which one produced an averaged record for the
charge induced by the triggering segment in  each preamplifier \cite{Radford:private}.  In
order to develop superpulses it was first necessary to perform single segment
cuts for each segment and cut for events for which the entire $\gamma$-ray
energy is deposited in the triggered segment.  The events' traces
were then averaged together to form a superpulse for each individual
channel (see Fig. \ref{superpfig}).  The superpulses were then digitally shaped using the offline
trapezoidal filter and the average energy value was calculated.  To
acquire the cross-talk coefficients superpulses were constructed for
several $\gamma$-ray energies providing an energy dependent response curve
for each individual channel, which was fit using a two parameter linear
function.  The slope determined by the fit is the cross-talk
coefficient.  The difference between the two methods is that in the
superpulse method, cuts are applied around the full peak $\gamma$-ray
energy and not the cross-talk band itself.

From these two methods it is possible to determine seventy-two
cross-talk coefficients of the 12$\times$8 cross-talk matrix, where the
remaining twenty-four are folded in with the calibration.  The values
obtained (See Table \ref{tab:intx2} in Appendix A) using both methods are acceptable and reflect the current
operating status of SEGA.  Most of the coefficients were less than
1\%, which is consistent with cross-talk values reported for other, segmented detectors \cite{Descovich:2005p109,Bruyneel:2009p93}.  However, there was significant
cross-talk between the two central contacts  whenever there was an
event in the upper C1 segment of the detector.  This is expected from the geometry
of the inner contacts and does not pose a significant problem for signal analysis.

\begin{figure}[H]
 \centering \includegraphics[width=1.0\textwidth, angle=0]{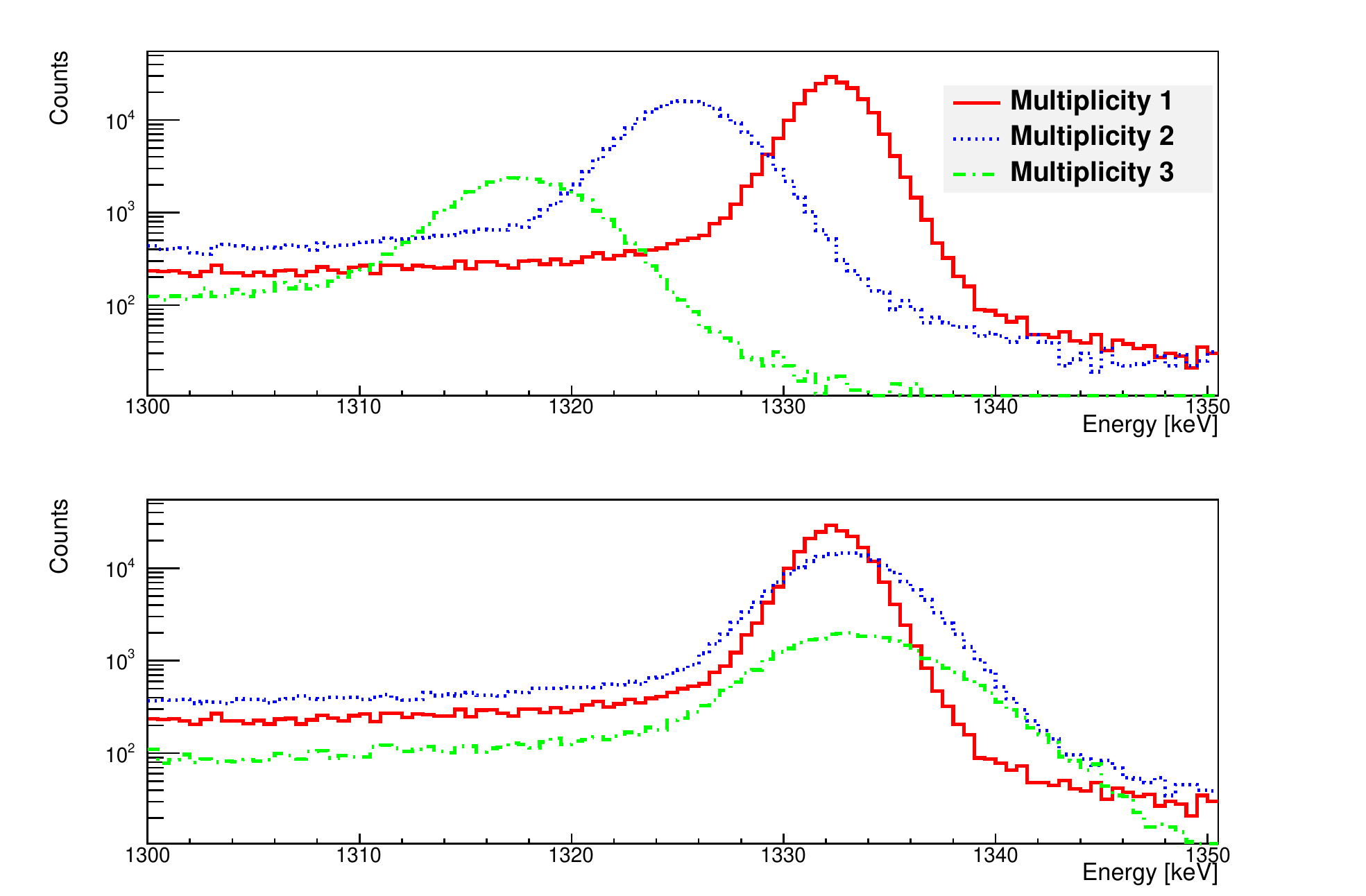}
 \caption{Total energy determined by summing the outer S channels (S1-S6) energies before (top) and after (bottom) the cross-talk correction was applied for multiplicities 1, 2, and 3.}
 \label{fig:xtalkcomp}
\end{figure}

An example of applying the cross-talk correction is shown in Fig. \ref{fig:xtalkcomp} where we have summed the outer S channels' (S1-S6) energies.  Without applying the cross-talk correction 
we see similar cross-talk effects as the AGATA collaboration \cite{Bruyneel:2009p93} where the photopeak's FWHM increases and the peak's position is shifted lower in energy as the 
multiplicity increases.  Using the cross-talk coefficients listed in Table \ref{tab:intx2} in Appendix A, we are able to realign and reconstruct the photopeaks.

\subsection{Electronic Noise}

Noise contributions to the energy resolution (expressed as the FWHM of the $\gamma$-ray full energy)
are specified in terms of equivalent noise charge (ENC in units of $e$). 
ENC can be converted to energy resolution by the following \cite{Knoll:book}
\begin{equation}
FWHM(eV)=2.35 \cdot 2.96(\frac{eV}{e}) \cdot ENC(e).
\end{equation}
As shown by \cite{Bertuccio:2009p62} the ENC introduced due to the detector-preamplifier
noise sources can be represented as 
\begin{equation}
ENC^{2}(e^{2}_{rms}) = \frac{h_{1}}{\tau} + h_{2} + h_{3} \tau
\end{equation}
\begin{equation}
h_{1}(e^{2}\mu s) = a C_{tot}^{2} A_{1}
\end{equation}
\begin{equation}
h_{2}(e^{2}) = \left(2 \pi a_{f}  C_{tot}^{2} \frac{b_{f}}{2 \pi}\right) A_{2}
\end{equation}
\begin{equation}
h_{3}(\frac{e^{2}}{\mu s}) = b A_{3},
\end{equation}
where $\tau$ is the peaking time of the shaping network, $a$ is the series white
noise, $b$ is the parallel white noise, $a_{f}$ and $b_{f}$ are the non-white series
and parallel noise,  $C_{tot}$ is the total input capacitance, and $A_{1}$, $A_{2}$,
and $A_{3}$ are weighting factors
dependent upon the shaping network that can be found in \cite{GATTI:1990p133}\footnote{The $A_{1}$, $A_{2}$,
and $A_{3}$ values for a trapezoidal filter are 2.00, 1.38, and 1.67 respectively.} (See also Appendix B).

\begin{figure}[H]
\begin{center}
\includegraphics
[width=1.0\textwidth, angle=0]{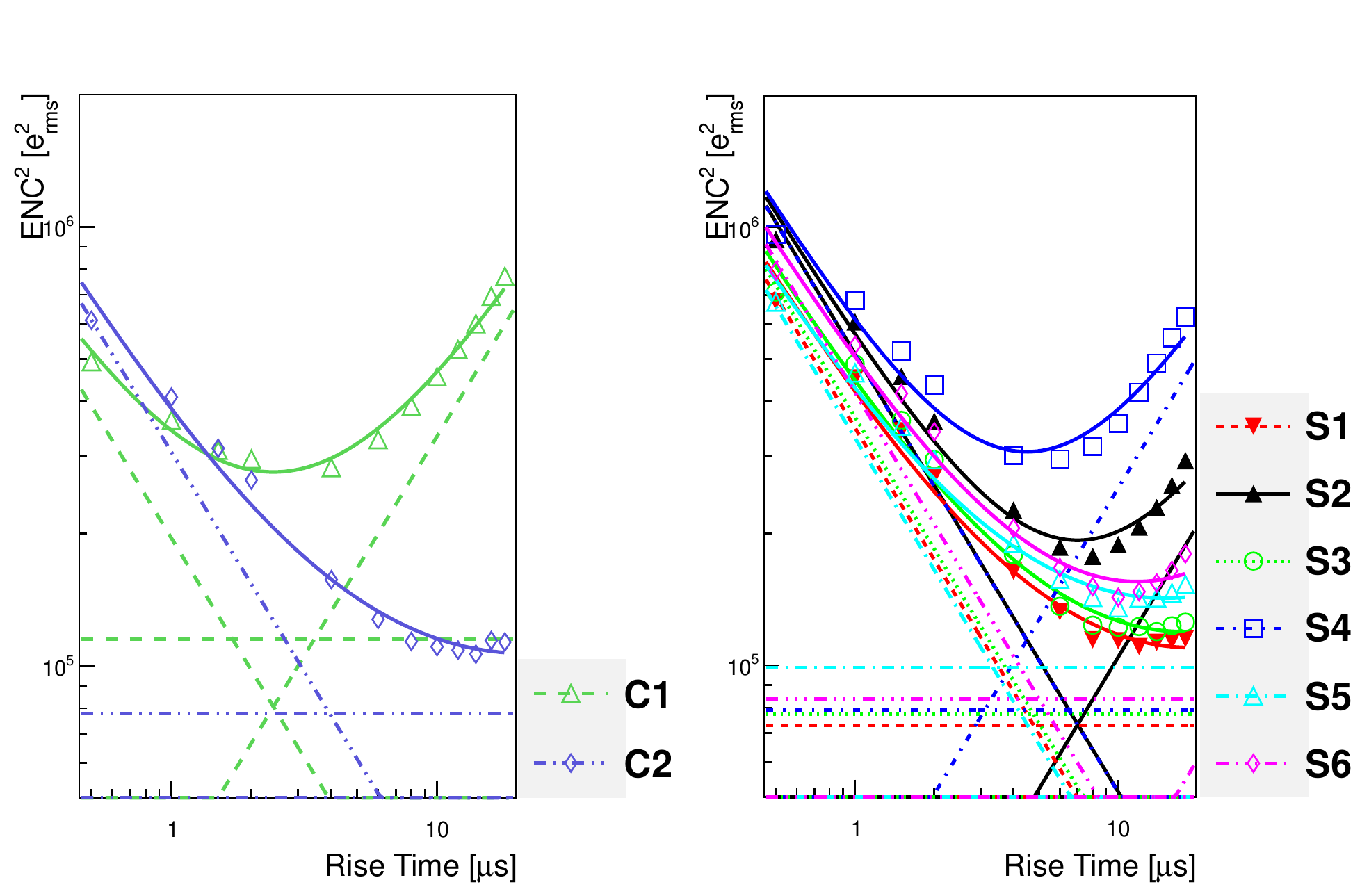}
\caption{For each readout contact, we depict ENC$^{2}$ {\it versus} rise-time curves along with the fitted contributions for white series, white parallel and non-white noise (see coded legend at right of figure; noise components are represented by dashed/dotted lines with exception to S2 being solid).  The peaking time for a trapezoidal filter shaping network is its rise-time \cite{GATTI:1990p133}.}
\label{encst}
\end{center}
\end{figure}

Our objective was to quantify the noise sources by varying the peaking time of our
shaping network.  Typically a pulser is used to decouple the additional contributing
factors to the measured peak width, but since SEGA currently is not equipped with
a pulser we placed a $^{60}$Co  source above
the crystal and recorded data with very long traces (40 $\mu$s).  Having 40 $\mu$s
traces allowed us to apply an offline digital trapezoidal filter outlined in \cite{Tan:2004p22}, 
which is the same filter implemented in the XIA module, while varying the
shaping parameters, eliminating the need for multiple data sets for which the
shaping parameters were varied from run to run.  FWHM values were measured by
fitting the  1332 keV photopeak in the trapezoidal filter output spectra using a
Gaussian function with a linear background.  To obtain only the electronic noise
contribution to the photopeak width all other components, including charge creation
statistics and charge collection efficiency were removed using the coefficients
obtained from the resolution versus energy model presented in the following section.  The ENC$^{2}$ {\it versus} $\tau$
curves in Fig. \ref{encst} provide some insight to how the three noise sources contribute
individually 
to the peak width.

Of the three noise contributions it is evident that the parallel noise exhibits the
most non-uniformity from channel to channel and is the dominant noise source for a few channels at rise times
longer than 7 $\mu$s.
It is also clear that the current system is limited by the series noise component and significant improvement 
of this component is expected with the transfer to an optimized cryostat.  A more detailed treatment of the parallel 
noise component is included in Appendix B.

\subsection{Resolution}

Due to the risk associated with each installation of a new set of contacts, a
significant effort was made to evaluate the performance of the 
SEGA detector in its temporary cryostat.  At the current noise levels microphonics 
were found to be negligible, but the overall electronic noise (due to a combination of grounding and 
shielding issues) was significant and not optimal. 

\begin{figure}[H]
 \centering
  \includegraphics[width=1.0\textwidth, angle=0]{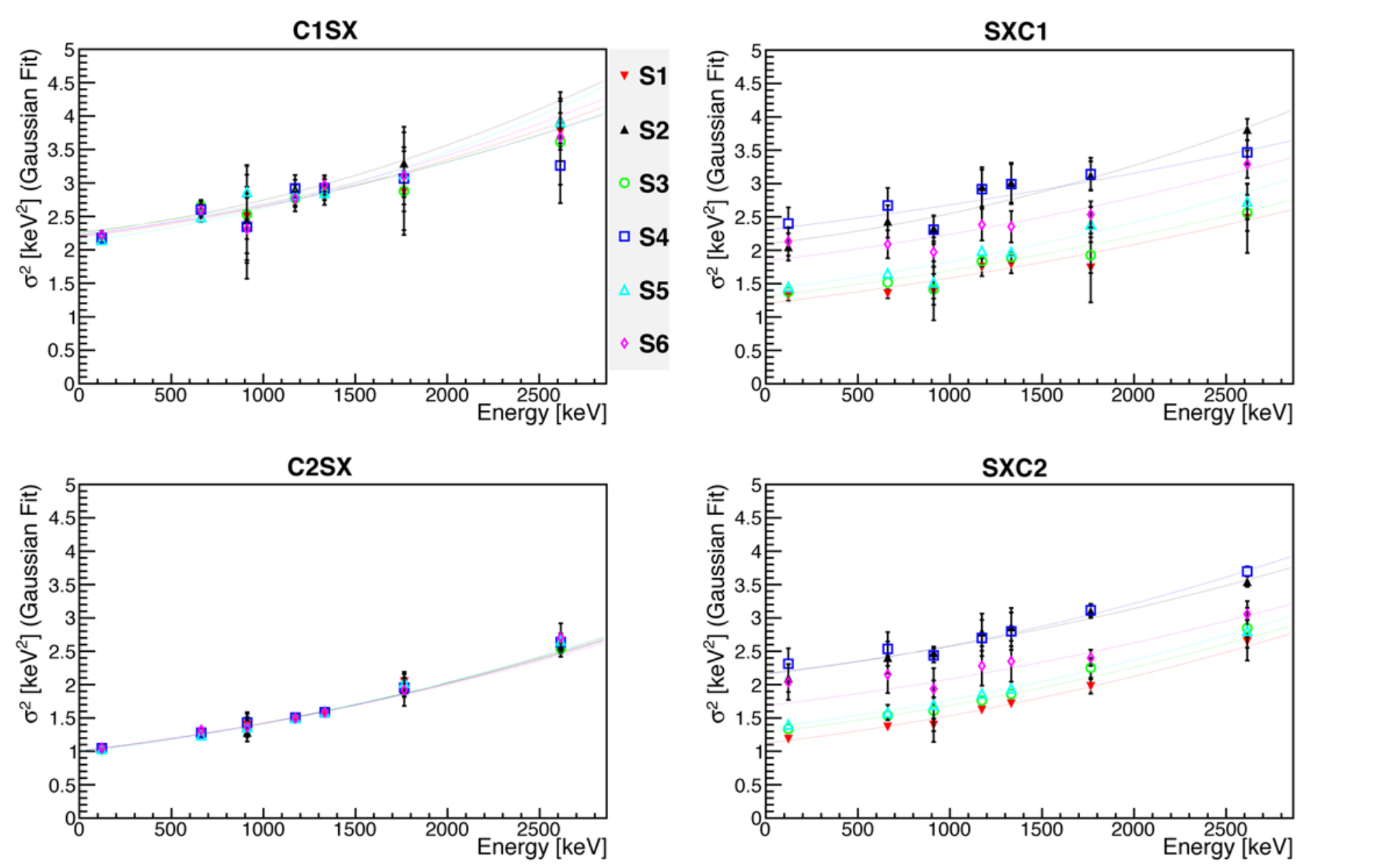}
 \caption{$\sigma$$^{2}$ {\it versus} energy for Gaussian fits to the full-energy peak for all physical segments.  The solid curves are the result of fits for a quadratic model for $\sigma$$^{2}$ as a function of energy. The sources used for this study are listed in Table \ref{tab:ressource}.}
 \label{s2eplot}
\end{figure}

The resolution was measured using a conventional spectroscopic amplifier with
several different shaping times, the on-board energy filter of 
the XIA DGF-4C module and an offline implementation of a trapezoidal filter.
For the data presented in Fig. \ref{s2eplot} and Tables \ref{tab:noiseandfwhm} and \ref{tab:fw2} the trapezoidal filter rise and gap times 
were set to 6.0 $\mu$s and 1.2 $\mu$s, respectively.
The sources used for the resolution measurements along with the measured rate, livetime,
and $\gamma$-ray peaks used for each source are listed in Table \ref{tab:ressource}.

\begin{table}[H]
\begin{center}
 \begin{tabular}{cccc}
   \hline
   Source &  Measured Rate [counts/s] & Livetime \% & Peaks  [keV]\\ \hline
   $^{57}$Co & 577 &  25.5 & 122 \\
   $^{60}$Co & 1880 & 6.6 & 1173, 1332 \\ 
   $^{137}$Cs & 353 & 41.2 & 662 \\ 
   Th & 86 & 65.1 & 911, 1764, 2615\\ \hline
    \end{tabular}
\end{center}
\caption{ Sources used for resolution study.}
 \label{tab:ressource}
\end{table}

Since each physical segment is unique, it was necessary to perform
resolution measurements of both read-outs (central and outer) for each physical
segment.  Full width at half maximum (FWHM), full width at tenth maximum (FWTM), and full width at fifty maximum (FWFM)
values were determined by fitting photopeaks using both a Gaussian function and modified Gaussian function.  Results of these measurements
are shown in Table \ref{tab:fw2} in Appendix C.

The resolution measurements were interpreted to quantify the factors that contribute
to the photopeak width and included:  the inherent statistical spread in the number
of charge carriers (W$_{D}$), variations in the charge-collection efficiency (W$_{X}$), and
contributions from electronic noise (W$_{E}$).  Since each of these three terms has a Gaussian
distribution, they add in quadrature and the photopeak width can be composed of
these terms as follows
\begin{equation}
 W_{T}^{2} = W_{D}^{2}+W_{X}^{2}+W_{E}^{2}.
 \end{equation}
  \begin{equation}
 W_{T}^{2} = aE+bE^{2}+c.
 \end{equation}
 
The charge-creation process factor is inherent to the material and is given by
\begin{equation}
W_{D}^{2}=(2.35)^{2}F \epsilon E
\end{equation}
where F is the Fano factor, $\epsilon$ is the
average energy needed to create an electron-hole pair, and E is the incident
$\gamma$-ray energy  \cite{Knoll:book}.  A range of Fano factor values exist in the literature, but for the data presented in Fig. \ref{s2eplot} 
and Table \ref{tab:noiseandfwhm} the Fano factor value of 0.11 was used \cite{Drummond:1971p91}.
The completeness of charge-collection term, W$_{X}^{2}$ (the term proportional to $b$ in Eq. 7), is
quadratic in terms of the incident $\gamma$-ray energy and is a property of the detector
geometry and impurity profile.  As expected, charge trapping varies throughout the detector
volume and it is exacerbated in regions where the electric field is small.  The
influences from charge trapping are especially apparent in the C1 physical segment
where the electric fields are low.  The
final contribution to the peak width is the electronic noise of the system which
does not vary with the incident $\gamma$-ray energy (See Table \ref{tab:noiseandfwhm} for fit values of the W$_{X}^{2}$ and W$_{E}^{2}$ terms).

Given this model of how the
resolution should vary with incident $\gamma$-ray energy, the evaluated peak widths were
fit with a second order polynomial as a function of $\gamma$-ray energy, which allowed
us to assess the contributing factors to the photopeak width.  In efforts to establish a fixed Fano factor value for all 
channels, values from 0.06 to 0.13 were tested.  It was determined that varying the Fano factor value within this range only changed the measured
and projected FWHM values at 2040 keV by less than 0.9\% and 2.1\% respectively.

During the evaluation of the resolution data it was found that three of the outer
channels' resolution performance exhibited higher rate sensitivity.  The
source of this behavior is currently unclear, but to account for this 10\% error
bars
were placed on the higher-rate data for these channels.  Following this, the error
bars were increased by a calculated scale factor in order to reduce the $\chi^{2}$/dof
to one.  From these fits the coefficients provide a characterization of
the contributing factors to the photopeak width.  The extrapolated electronic noise and charge-collection terms for each 
channel are listed in Table \ref{tab:noiseandfwhm}.

\begin{table}[H]
\begin{center}
\scalebox{0.75}{
 \begin{tabular}{ccccc}
   \hline
   \multirow{2}{*}{Segment} & W$_{E}^{2}$ (FWHM$^{2}$) & W$_{X}^{2}$ (FWHM$^{2}$) & FWHM (2040 keV) & FWHM (2040 keV)\\
   & [keV$^{2}$] & [keV$^{2}$] & Measured [keV] & Projected [keV]\\ \hline
C1S1 & 12.1 $\pm$ 0.2 & 2.88 $\pm$ 0.70 & 4.32 $\pm$ 0.08 & 2.75 $\pm$ 0.14\\ 
C1S2 & 12.4 $\pm$ 0.3 & 3.81 $\pm$ 1.10 & 4.46 $\pm$ 0.12 & 2.91 $\pm$ 0.19\\ 
C1S3 & 12.5 $\pm$ 0.4 & 2.35 $\pm$ 1.60 & 4.30 $\pm$ 0.19 & 2.65 $\pm$ 0.32\\ 
C1S4 & 12.3 $\pm$ 0.4 & 2.50 $\pm$ 1.50 & 4.30 $\pm$ 0.18 & 2.68 $\pm$ 0.30\\ 
C1S5 & 11.8 $\pm$ 0.2 & 3.90 $\pm$ 0.70 & 4.40 $\pm$ 0.08 & 2.93 $\pm$ 0.13\\ 
C1S6 & 12.2 $\pm$ 0.2 & 3.16 $\pm$ 0.84 & 4.36 $\pm$ 0.10 & 2.80 $\pm$ 0.16\\ 
C2S1 & 5.54 $\pm$ 0.08 & 2.09 $\pm$ 0.26 & 3.36 $\pm$ 0.04 & 2.60 $\pm$ 0.05\\ 
C2S2 & 5.53 $\pm$ 0.05 & 2.18 $\pm$ 0.16 & 3.37 $\pm$ 0.03 & 2.62 $\pm$ 0.03\\ 
C2S3 & 5.54 $\pm$ 0.07 & 2.12 $\pm$ 0.24 & 3.37 $\pm$ 0.04 & 2.61 $\pm$ 0.05\\ 
C2S4 & 5.55 $\pm$ 0.08 & 2.06 $\pm$ 0.24 & 3.36 $\pm$ 0.04 & 2.59 $\pm$ 0.05\\ 
C2S5 & 5.44 $\pm$ 0.06 & 2.29 $\pm$ 0.20 & 3.38 $\pm$ 0.03 & 2.64 $\pm$ 0.04\\ 
C2S6 & 5.58 $\pm$ 0.12 & 1.95 $\pm$ 0.37 & 3.35 $\pm$ 0.06 & 2.57 $\pm$ 0.08\\ 
S1C1 & 6.63 $\pm$ 0.29 & 1.34 $\pm$ 1.20 & 3.41 $\pm$ 0.18 & 2.45 $\pm$ 0.26\\ 
S2C1 & 11.6 $\pm$ 0.6 & 3.02 $\pm$ 0.71 & 4.27 $\pm$ 0.11 & 2.77 $\pm$ 0.21\\ 
S3C1 & 7.23 $\pm$ 0.15 & 1.44 $\pm$ 0.66 & 3.51 $\pm$ 0.10 & 2.47 $\pm$ 0.14\\ 
S4C1 & 12.7 $\pm$ 0.7 & 1.14 $\pm$ 0.80 & 4.19 $\pm$ 0.13 & 2.41 $\pm$ 0.27\\ 
S5C1 & 7.80 $\pm$ 0.13 & 2.05 $\pm$ 0.53 & 3.68 $\pm$ 0.08 & 2.59 $\pm$ 0.11\\ 
S6C1 & 10.2 $\pm$ 0.6 & 1.73 $\pm$ 0.84 & 3.95 $\pm$ 0.13 & 2.53 $\pm$ 0.24\\ 
S1C2 & 6.19 $\pm$ 0.07 & 2.02 $\pm$ 0.20 & 3.45 $\pm$ 0.03 & 2.59 $\pm$ 0.04\\ 
S2C2 & 12.0 $\pm$ 0.4 & 1.85 $\pm$ 0.40 & 4.18 $\pm$ 0.07 & 2.55 $\pm$ 0.14\\ 
S3C2 & 7.08 $\pm$ 0.08 & 1.89 $\pm$ 0.25 & 3.56 $\pm$ 0.04 & 2.56 $\pm$ 0.05\\ 
S4C2 & 11.9 $\pm$ 0.5 & 2.36 $\pm$ 0.42 & 4.23 $\pm$ 0.07 & 2.65 $\pm$ 0.15\\ 
S5C2 & 7.47 $\pm$ 0.30 & 2.10 $\pm$ 0.90 & 3.64 $\pm$ 0.13 & 2.60 $\pm$ 0.19\\ 
S6C2 & 9.31 $\pm$ 0.55 & 1.67 $\pm$ 0.51 & 3.83 $\pm$ 0.10 & 2.52 $\pm$ 0.18\\ \hline
\end{tabular}
}
\end{center}
\caption{Measured electronic noise and charge-collection contributions to the photopeak width along with the measured and projected FWHM values at 2040 keV for 6 $\mu$s rise time.  
The projected values assume 1.0 keV FWHM for the electronic noise contribution and the charge-collection contribution was calculated at 2040 keV.  Errors are statistical uncertainties from Gaussian fits to the photopeaks and 
the errors from the extracted values for the W$_{E}^{2}$ parameter. }
\label{tab:noiseandfwhm}
\end{table}

\section{Discussion and Analysis}
\label{}

The performance of this prototype detector must be evaluated in terms of how it would
perform in a double beta-decay experiment.  Most of the performance 
specifications are already within reasonable limits.  The crystal 
impurity concentrations were near the limits of what is typically considered useful 
detector material at ORTEC; the bias potential was quite close to ORTEC's practical 
working limit of 5 kV.  It is assumed that this applies specifically to the crystal pulled for 
SEGA, and not to $^{76}$Ge-enriched material in general.

The availability of a strictly limited amount of $^{76}$Ge (as opposed to full-scale, enriched Ge detector 
development) restricted the amount of processing and the achieved operational parameters of this detector, and resulted in accepting rather high leakage currents and the 
lack of bulletization.
The leakage currents resulted in one physical segment S4, exhibiting parallel noise which limited
its resolution for shaping times longer than 7 $\mu$s (Sec. 4.7 and Appendix B).  The lack of bulletization resulted
in poorer charge collection characteristics for the C1-coupled segments relative to the C2-coupled segments, as
evidenced by larger low-energy tails in the C1-coupled segments (Sec. 4.1 and Appendix C) and the larger W$_{X}^{2}$
parameter in the behavior of the resolution as a function of energy (Sec 4.1 and Appendix C).  Clearly, both leakage currents
and lack of bulletization had some impact on the resolution, but do not seem to a be a serious, intrinsic impediment for 
this geometry.

On the other hand, the only significant issue which emerged for this detector is its marginal achievement of the resolution specification (4 keV FWHM at 2039 keV).
Even allowing for the degradation in performance due to leakage currents and lack of bulletization, Table \ref{tab:noiseandfwhm} shows
that all but two physical segments have at least one readout which meets the nominal {\sc Majorana} resolution goal and are within the detector resolution range for the Heidelberg-Moscow experiment \cite{Klapdor:2001:p148}\footnote{Table \ref{tab:fw2} in Appendix C shows all channels have a FWHM at 1332 keV of at least 0.6 keV greater than the IGEX detectors \cite{Aalseth:1999:p2109}.}.  
These segments, C1S2 and C1S4, should also reach the {\sc Majorana} resolution goal if somewhat shorter, optimal rise time values were 
used for channel C1, so we conclude all physical segments have at least one channel satisfying the minimal requirement for the 
{\sc Majorana} project.  We also note that the non-enriched prototype PT6x2 with cold FETs had a much smaller FWHM measured for the C1 channel at 122 keV (0.9 keV) and
generally improved resolution relative to SEGA at 1332 keV (all S channel FWHM values at 1332 keV were less than 3.2 keV and the C2 channel had a FWHM value at 1332 keV of 2.4 keV) \cite{Sangsingkeow:2003p34}, indicating that detectors with the
PT6x2 geometry should, in general, meet the {\sc Majorana} resolution goal.

The performance of SEGA with cold FETs can be projected.  
From the resolution versus energy scan, it is clear that the energy resolution is limited primarily by electronic 
noise.  The shaping time scans show that for the S4 and possibly C1 channels, the electronic noise  
can be attributed primarily to the leakage current.  All segments however, also exhibit series 
noise contributions which limit the resolution at shaping times less than 10 $\mu$s, consistent with 
the attribution of large electronic noise due to the shielding and grounding issues in the temporary cryostat.

The noise contribution for a standard detector with properly grounded, shielded, and
cooled FETs should be below roughly 0.7 keV FWHM.  ORTEC estimates 1.0 keV may be
appropriate for warm FETs.  Comparing the parameters from the $\sigma$$^{2}$ fits,
one can develop an understanding of how large the noise contributions are due to the
warm FETs and the temporary cryostat.  In every case, larger noise contributions are
observed than expected in an optimal system.
The C1 and C2 channels currently have
noise widths (FWHM) around 3.3 keV and 2.3 keV, respectively.  The outer channels
S1-S6 have a large range in noise widths of 2.5-3.3 keV.  By transferring the
crystal to an improved cryostat, the series noise
contribution should be significantly reduced, and reach the targeted resolution of below 4 keV FWHM at 2039 keV
for all channels. Table \ref{tab:noiseandfwhm} provides the current measured FWHM at 2040 keV
along with the projected FWHM if 1.0 keV electronic noise is assumed.

For the SEGA PT6x2 detector, the assumption that 1.0 keV
electronic noise is achievable for the channels where leakage current is the
dominant electronic noise source may be overly optimistic. Even for these channels,
however, only about a 5\% improvement in the current resolution is required to
reach the {\sc Majorana} resolution goal.  This should certainly be achievable
with optimized grounding, shielding, shaping parameters and cold FETs.

The fact that the leakage current flows predominately through the internal C1 contact along with the leakage current measurements of
the non-enriched prototype PT6x2 detector could suggest
that the leakage current behavior in SEGA may be a characteristic of the PT6x2 
fabrication procedure.  
As pointed out in Sec. 4.3, the etching of SEGA to reduce leakage current
was curtailed to preserve enriched material.  Given the reasonable resolution performance of SEGA, it is expected that it will be possible 
to consistently meet acceptable leakage current limits if manufacturers operate in a ``production" mode
where stringent leakage current specifications can be applied to enriched crystals.
In this case, crystals can be etched until leakage currents fall below a reasonable
upper limit or the entire crystal can be returned to raw germanium to be reprocessed. This
conclusion is supported by the performance of the existing PT6x2 prototype, which demonstrated
acceptable resolution specifications.

\section{Conclusion}
\label{}

We have reported on the successful fabrication, for the first time, of an isotopically enriched segmented N-type germanium detector.
From the studies presented here and background rejection studies performed,
in part, at the HIGS facility at the Triangle Universities Nuclear Laboratory \cite{SEGAPSA}, we believe 
the PT6x2 geometry is a potentially viable prototype for an N-type germanium 
detector experiment to search for neutrinoless double beta-decay \cite{Abt:2007p479,Abt:2007p332,Guiseppe:2008}.

\section*{Acknowledgments}
We would like to thank the TUNL technical staff for their many contributions towards
this project.  We would also like to acknowledge helpful discussion and feedback from A. W. Poon and R. D. Martin in the process of preparing this manuscript.
This work was supported in part by the 
Office of Science of the US Department of Energy (DOE) under grant number DE-FG02-97ER41042.

\appendix
\section{Cross-talk}

The 12$\times$8 integral cross-talk matrix listed in Table \ref{tab:intx2} was generated using the 
superpulse method described in the cross-talk section (the cross-talk matrix from the band method
is in agreement with Table \ref{tab:intx2} and excluded to avoid redundancy).  Seventy-two of the ninety-six coefficients
were determined from the fits to the response curves while the remaining twenty-four are
folded in with the calibration coefficients and are represented by unity.

\begin{table}[H]
\begin{center}
\scalebox{0.6}{
 \begin{tabular}{ccccccccc}
   \hline
         & S1 & S2 & S3 & S4 & S5 & S6 & C1 & C2 \\ \hline
C1S1 & 1 & -0.007 $\pm$ 0.001 & -0.009 $\pm$ 0.001 & -0.009 $\pm$ 0.001 & -0.009 $\pm$ 0.001 & -0.007 $\pm$ 0.001 & 1 & 0.033 $\pm$ 0.001\\ 
C1S2 & -0.008 $\pm$ 0.002 & 1 & -0.007 $\pm$ 0.001 & -0.009 $\pm$ 0.001 & -0.009 $\pm$ 0.001 & -0.009 $\pm$ 0.001 & 1 & 0.034 $\pm$ 0.001\\ 
C1S3 & -0.007 $\pm$ 0.002 & -0.009 $\pm$ 0.001 & 1 & -0.006 $\pm$ 0.002 & -0.009 $\pm$ 0.001 & -0.009 $\pm$ 0.001 & 1 & 0.034 $\pm$ 0.001\\ 
C1S4 & -0.009 $\pm$ 0.001 & -0.009 $\pm$ 0.001 & -0.007 $\pm$ 0.001 & 1 & -0.010 $\pm$ 0.001 & -0.009 $\pm$ 0.001 & 1 & 0.035 $\pm$ 0.002\\ 
C1S5 & -0.009 $\pm$ 0.001 & -0.009 $\pm$ 0.001 & -0.009 $\pm$ 0.001 & -0.005 $\pm$ 0.002 & 1 & -0.007 $\pm$ 0.001 & 1 & 0.039 $\pm$ 0.001\\ 
C1S6 & -0.009 $\pm$ 0.001 & -0.008 $\pm$ 0.001 & -0.010 $\pm$ 0.001 & -0.009 $\pm$ 0.001 & -0.003 $\pm$ 0.001 & 1 & 1 & 0.078 $\pm$ 0.003\\ 
C2S1 & 1 & -0.005 $\pm$ 0.001 & -0.007 $\pm$ 0.001 & -0.004 $\pm$ 0.001 & -0.007 $\pm$ 0.001 & -0.006 $\pm$ 0.001 & 0.010 $\pm$ 0.001 & 1\\ 
C2S2 & -0.007 $\pm$ 0.001 & 1 & -0.005 $\pm$ 0.001 & -0.005 $\pm$ 0.001 & -0.006 $\pm$ 0.001 & -0.007 $\pm$ 0.001 & 0.010 $\pm$ 0.001 & 1\\ 
C2S3 & -0.007 $\pm$ 0.001 & -0.006 $\pm$ 0.001 & 1 & -0.002 $\pm$ 0.002 & -0.006 $\pm$ 0.001 & -0.005 $\pm$ 0.001 & 0.010 $\pm$ 0.001 & 1\\ 
C2S4 & -0.007 $\pm$ 0.001 & -0.007 $\pm$ 0.001 & -0.005 $\pm$ 0.001 & 1 & -0.008 $\pm$ 0.001 & -0.007 $\pm$ 0.001 & 0.010 $\pm$ 0.001 & 1\\ 
C2S5 & -0.006 $\pm$ 0.001 & -0.007 $\pm$ 0.001 & -0.006 $\pm$ 0.001 & -0.002 $\pm$ 0.001 & 1 & -0.006 $\pm$ 0.001 & 0.010 $\pm$ 0.001 & 1\\ 
C2S6 & -0.007 $\pm$ 0.001 & -0.007 $\pm$ 0.001 & -0.006 $\pm$ 0.001 & -0.005 $\pm$ 0.001 & -0.006 $\pm$ 0.001 & 1 & 0.010 $\pm$ 0.001 & 1\\ \hline
 \end{tabular}
 }
\end{center}
\caption{Integral cross-talk coefficients.}
\label{tab:intx2}
\end{table}

\section{Electronic Noise}

Electronic noise coefficients from the fits to the ENC$^{2}$ {\it versus} rise-time curves are listed in Table \ref{tab:noisecoef}.
These coefficients characterize the three electronic noise components of the detector-preamplifier system, which include
series, non-white and parallel noise.

\begin{table}[H]
\begin{center}
 \begin{tabular}{cccc}
   \hline
   Segment & h1 [$e^{2}$$\mu$s] & h2 [$e^{2}$] & h3 [$e^{2}$/$\mu$s]  \\ \hline
S1 & 3.48e+05 $\pm$ 6e+03 & 5.77e+04 $\pm$ 3e+03 & 976 $\pm$ 3e+02\\
S2 & 5.14e+05 $\pm$ 9e+03 & 3.13e+04 $\pm$ 6e+03 & 1.04e+04 $\pm$ 5e+02\\ 
S3 & 3.68e+05 $\pm$ 7e+03 & 6.22e+04 $\pm$ 4e+03 & 1.19e+03 $\pm$ 3e+02\\ 
S4 & 5.12e+05 $\pm$ 1e+04 & 6.39e+04 $\pm$ 7e+03 & 2.54e+04 $\pm$ 7e+02\\ 
S5 & 3.3e+05 $\pm$ 7e+03 & 8.35e+04 $\pm$ 4e+03 & 1.44e+03 $\pm$ 3e+02\\ 
S6 & 4.19e+05 $\pm$ 8e+03 & 6.87e+04 $\pm$ 5e+03 & 3.04e+03 $\pm$ 4e+02\\ 
C1 & 1.95e+05 $\pm$ 7e+03 & 9.98e+04 $\pm$ 7e+03 & 3.33e+04 $\pm$ 8e+02\\ 
C2 & 3.07e+05 $\pm$ 6e+03 & 6.27e+04 $\pm$ 4e+03 & 680 $\pm$ 3e+02\\ \hline
 \end{tabular}
\end{center}
\caption{Electronic noise coefficients from ENC$^{2}$ {\it versus} rise-time curves.}
\label{tab:noisecoef}
\end{table}

\begin{table}[H]
\begin{center}
 \begin{tabular}{ccc}
   \hline
   Segment & Feedback R [pA] &   $q$$h_{3}$/$A_{3}$  [pA]\\ \hline
S1 & 4 & 93.5 \\ 
S2 & 5 & 997 \\ 
S3 & 234 & 114 \\ 
S4 & 2500 & 2430 \\ 
S5 & 100 & 138 \\ 
S6 & 90 & 291 \\ 
C1 & N/A & 3190 \\ 
C2 & N/A & 65.1 \\ \hline
 \end{tabular}
\end{center}
\caption{Leakage current as measured across the feedback resistor with a voltmeter and calculated
from the h3 coefficient from the ENC$^{2}$ {\it versus} rise-time curves.  The feedback resistor and h3 coefficient measurements were done
in October, 2009 and December, 2009 respectively}
\label{tab:noisecoef}
\end{table}

Detailed measurements with 40 $\mu$s traces were conducted in 2009, motivating the reinvestigation of the leakage currents
measured in 2003.  The 2009 measurements are in qualitative agreement with the 2003 data (S4 dominate source of leakage current) 
and indicates, for most channels, the parallel noise is roughly consistent with expectations.  We do not understand the reason our leakage current increased, or 
why the values calculated for some channels are inconsistent with the feedback resistor measurements.  Because the detector was warmed, pumped, and recooled
several times between 2003 and 2009, changes in surface conditions on the contacts may be partially responsible for the observed increase in leakage currents.

\section{Resolution}

FWHM values at 1332 keV using both a Gaussian functional fit and a modified Gaussian functional fit \cite{Debertin:1988p164} are listed in Table \ref{tab:fw2}.  
For all but two segments there is at least one channel with a measured value 
which reaches the targeted resolution goal of 4 keV FWHM at 2039 keV and improvement is expected once several grounding and shielding issues are resolved.
It should also be noted that by optimizing the rise-time values for the trapezoidal filter all segments have at least one channel
reaching the {\sc Majorana} specification.  Also listed are the FWTM/FWHM and FWFM/FWHM ratios for the modified Gaussian functional fit, along with the 
integrated number of counts in the low-energy tail as a ratio to the number of counts in the photopeak.  The deviations from
a Gaussian functional fit for the FWFM/FWHM and integrated number of counts in the low-energy tail both suggest charge-collection issues are present for the SXC1 channels.

\begin{table}[H]
\begin{center}
\scalebox{0.6}{
 \begin{tabular}{cccccc}
   \hline
   \multirow{2}{*}{Segment} & FWHM at 1332 keV & FWHM at 1332 keV  & FWTM/FWHM & FWFM/FWHM & Low-Energy \\
   & (Gaussian Fit) [keV]  & (Modified Gaussian Fit) [keV]  & (Modified Gaussian Fit) & (Modified Gaussian Fit)
 & Tail\\ \hline
C1S1 & 4.08 $\pm$ 0.04 & 3.98 $\pm$ 0.10 & 1.85 $\pm$ 0.06 & 2.71 $\pm$ 0.30 & 0.021 $\pm$ 0.002\\ 
C1S2 & 4.12 $\pm$ 0.04 & 4.04 $\pm$ 0.10 & 1.86 $\pm$ 0.07 & 2.73 $\pm$ 0.30 & 0.007 $\pm$ 0.007\\ 
C1S3 & 4.03 $\pm$ 0.04 & 3.96 $\pm$ 0.10 & 1.83 $\pm$ 0.05 & 2.50 $\pm$ 0.10 & 0.009 $\pm$ 0.002\\ 
C1S4 & 4.08 $\pm$ 0.04 & 4.04 $\pm$ 0.10 & 1.83 $\pm$ 0.05 & 2.48 $\pm$ 0.09 & 0.019 $\pm$ 0.003\\ 
C1S5 & 4.02 $\pm$ 0.04 & 3.96 $\pm$ 0.10 & 1.84 $\pm$ 0.05 & 2.52 $\pm$ 0.09 & 0.010 $\pm$ 0.002\\ 
C1S6 & 4.09 $\pm$ 0.04 & 3.93 $\pm$ 0.10 & 1.86 $\pm$ 0.07 & 2.68 $\pm$ 0.20 & 0.013 $\pm$ 0.002\\ 
C2S1 & 2.99 $\pm$ 0.01 & 2.96 $\pm$ 0.02 & 1.83 $\pm$ 0.01 & 2.41 $\pm$ 0.04 & 0.007 $\pm$ 0.001\\ 
C2S2 & 3.02 $\pm$ 0.01 & 3.00 $\pm$ 0.02 & 1.83 $\pm$ 0.02 & 2.48 $\pm$ 0.02 & 0.023 $\pm$ 0.001\\ 
C2S3 & 3.01 $\pm$ 0.01 & 3.01 $\pm$ 0.04 & 1.83 $\pm$ 0.03 & 2.38 $\pm$ 0.03 & 0.006 $\pm$ 0.001\\ 
C2S4 & 2.99 $\pm$ 0.01 & 2.99 $\pm$ 0.02 & 1.82 $\pm$ 0.01 & 2.48 $\pm$ 0.01 & 0.019 $\pm$ 0.001\\ 
C2S5 & 2.99 $\pm$ 0.01 & 2.98 $\pm$ 0.02 & 1.83 $\pm$ 0.02 & 2.41 $\pm$ 0.03 & 0.005 $\pm$ 0.001\\ 
C2S6 & 2.97 $\pm$ 0.01 & 2.95 $\pm$ 0.02 & 1.83 $\pm$ 0.01 & 2.42 $\pm$ 0.02 & 0.007 $\pm$ 0.001\\ 
S1C1 & 3.29 $\pm$ 0.03 & 3.13 $\pm$ 0.07 & 1.90 $\pm$ 0.05 & 4.26 $\pm$ 0.20 & 0.064 $\pm$ 0.003\\ 
S2C1 & 4.27 $\pm$ 0.03 & 4.10 $\pm$ 0.08 & 1.89 $\pm$ 0.05 & 3.56 $\pm$ 0.10 & 0.042 $\pm$ 0.002\\ 
S3C1 & 3.39 $\pm$ 0.03 & 3.32 $\pm$ 0.08 & 1.85 $\pm$ 0.05 & 3.19 $\pm$ 0.10 & 0.028 $\pm$ 0.002\\ 
S4C1 & 4.22 $\pm$ 0.03 & 4.13 $\pm$ 0.10 & 1.86 $\pm$ 0.05 & 3.35 $\pm$ 0.10 & 0.038 $\pm$ 0.002\\ 
S5C1 & 3.44 $\pm$ 0.03 & 3.33 $\pm$ 0.09 & 1.87 $\pm$ 0.06 & 3.64 $\pm$ 0.20 & 0.043 $\pm$ 0.002\\ 
S6C1 & 3.78 $\pm$ 0.03 & 3.63 $\pm$ 0.09 & 1.90 $\pm$ 0.05 & 3.97 $\pm$ 0.20 & 0.049 $\pm$ 0.003\\ 
S1C2 & 3.09 $\pm$ 0.01 & 3.06 $\pm$ 0.02 & 1.62 $\pm$ 0.20 & 2.38 $\pm$ 0.02 & 0.005 $\pm$ 0.002\\ 
S2C2 & 4.00 $\pm$ 0.01 & 4.00 $\pm$ 0.02 & 1.82 $\pm$ 0.01 & 2.40 $\pm$ 0.02 & 0.002 $\pm$ 0.001\\ 
S3C2 & 3.22 $\pm$ 0.01 & 3.21 $\pm$ 0.02 & 1.82 $\pm$ 0.01 & 2.41 $\pm$ 0.02 & 0.002 $\pm$ 0.001\\ 
S4C2 & 3.95 $\pm$ 0.01 & 3.99 $\pm$ 0.02 & 1.82 $\pm$ 0.01 & 2.42 $\pm$ 0.01 & 0.004 $\pm$ 0.001\\ 
S5C2 & 3.29 $\pm$ 0.01 & 3.28 $\pm$ 0.02 & 1.62 $\pm$ 0.20 & 2.36 $\pm$ 0.01 & 0.002 $\pm$ 0.001\\ 
S6C2 & 3.62 $\pm$ 0.01 & 3.53 $\pm$ 0.10 & 1.85 $\pm$ 0.08 & 2.41 $\pm$ 0.10 & 0.002 $\pm$ 0.002\\ \hline
 \end{tabular}
 }
\end{center}
\caption{FWHM values for each physical segment at 1332 keV from the
Gaussian functional fits and modified Gaussian functional fits.  Also listed are the FWTM/FWHM and FWFM/FWHM ratios
for the modified Gaussian functional fits, along with the integrated number of counts in the low-energy tail as a ratio to the number of counts in the photopeak.}
 \label{tab:fw2}
\end{table}




\begin{thebibliography}{00}
\bibliographystyle{elsarticle-num}
\bibitem{Aalseth:2004p148}C. E. Aalseth et al., Phys. Atom. Nucl., vol. 67, (2004), 2002
\bibitem{Schubert:2012p480}A. G. Schubert et al., AIP Conf. Proc. 1441, (2012), 480
\bibitem{Wilkerson:2012}J. F. Wilkerson et al., J. Phys. Conf. Ser. 375 042010, (2012)
\bibitem{Camilleri:2008p343}L. Camilleri, E. Lisi, J.F. Wilkerson, Annu. Rev. Nucl. Part. Sci., vol. 58, (2008), 343
\bibitem{III:2008p481}F. T. Avignone III, S. R. Elliott, and J. Engel, Rev. Mod. Phys., vol. 80, (2008), 481
\bibitem{Barabash:2010p162}A. S. Barabash, Phys. Atom. Nucl., vol. 73, (2010), 162
\bibitem{Barbeau:2007p53}P. S. Barbeau, J. I. Collar, and O. Tench, J. of Cosmol. and Astropart. Phys., vol. 9, (2007)
\bibitem{Phillips:2012}D. G. Phillips II et al., J. Phys. Conf. Ser. 381 012044, (2012)
\bibitem{Elliott:2006}S. R. Elliott et al., Nucl. Instr. and Meth. A558, (2006), 504
\bibitem{Deleplanque:1999p292}M.A. Deleplanque et al., Nucl. Instr. and Meth. A430, (1999), 292
\bibitem{Abt:2007p479} I. Abt et al., Nucl. Instr. and Meth. A570, (2007), 479
\bibitem{SEGAPSA}A. R. Young et al, Internal {\sc Majorana} document, (2007), Full publication of these results is in development
\bibitem{Descovich:2005p109}M. Descovich et al., Nucl. Instr. and Meth. A553, (2005), 535
\bibitem{Recchia:2009p111}F. Recchia et al., Nucl. Instr. and Meth. A604, (2009), 555
\bibitem{Abt:2007p474}I. Abt et al., Nucl. Instr. and Meth. A577, (2007), 474
\bibitem{Abt:2009}I. Abt et al., JINST 4 P11008, (2009)
\bibitem{itep:comp}The official ITEP homepage, URL:  http://www.itep.ru
\bibitem{igex:comp}The official IGEX homepage, URL:  http://www.nu.to.infn.it
\bibitem{dubna:comp}The official JINR homepage, URL:  http://www.jinr.ru
\bibitem{ortec:comp}The official ORTEC homepage, URL:  http://www.ortec-online.com
\bibitem{Sangsingkeow:2003p34}P. Sangsingkeow et al., Nucl. Instr. and Meth. A505, (2003), 183.  Please contact P. Sangsingkeow (Pat.Sangsingkeow@ametek.com) for more detailed information concerning the PT6x2 geometry.
\bibitem{Eberth:2001p228}J. Eberth et al., Prog. in Part. and Nucl. Phys., vol. 46, (2001), 389
\bibitem{Gamir:1997p113}P. Gamir, Ph. D. Dissertation, Cranfield University, (1997)
\bibitem{canberra:comp}The official Canberra homepage, URL:  http://www.canberra.com
\bibitem{xia:comp}The official XIA homepage, URL:  http://www.xia.com
\bibitem{Debertin:1988p164}K. Debertin and R.G. Helmer, $\gamma$ and X-Ray Spectrometry with Semiconductor Detectors,1988
\bibitem{ORTEC:private}ORTEC, Private Communication
\bibitem{Keyser:eff}R. M. Keyser and W. K. Hensley, IEEE Trans. Nucl. Sci., vol. 1, (2002), 275
\bibitem{Bruyneel:2009p93}B. Bruyneel et al., Nucl. Instr. and Meth. A608, (2009), 99
\bibitem{Boswell:2011p1212}M. Boswell et al., IEEE Trans. Nucl. Sci., vol. 58, (2011), 1212
\bibitem{Tan:2004p22}H. Tan et al., IEEE Trans. Nucl. Sci., vol. 51, (2004), 1541
\bibitem{Radford:private}D. Radford, Private Communication
\bibitem{Knoll:book}G. F. Knoll, Radiation Detection and Measurement, Wiley, 1979
\bibitem{Bertuccio:2009p62}G. Bertuccio and A. Pullia, Rev. of Sci. Instr., vol. 64, (2009), 3294
\bibitem{GATTI:1990p133}E. Gatti et al., Nucl. Instr. and Meth. A297, (1990), 467
\bibitem{Drummond:1971p91}W. E. Drummond, IEEE Trans. Nucl. Sci., vol. NS18, (1971), 91
\bibitem{Klapdor:2001:p148}H. V. Klapdor-Kleingrothaus, Eur. Phys. A 12, (2001), 148
\bibitem{Aalseth:1999:p2109}C. E. Aalseth et al., Phys. Ref. C, vol. 59, (1999), 2109
\bibitem{Abt:2007p332}I. Abt et al., Nucl. Instr. and Meth. A583, (2007), 332
\bibitem{Guiseppe:2008}V. E. Guiseppe et al., IEEE Nucl. Sci. Symp. Conf. Rec., (2008), 1793





\end{thebibliography}
\end{document}